\documentclass[prc,nofootinbib,superscriptaddress,preprintnumbers]{revtex4-2}

\pdfoutput=1

\usepackage{url}

\usepackage[utf8]{inputenc}
\usepackage{amsmath}
\usepackage{amssymb}
\usepackage{amsfonts}
\usepackage{xspace}
\usepackage{graphicx}
\usepackage{slashed}
\usepackage{bbold}
\usepackage{bbm}
\usepackage{physics}
\usepackage{empheq}
\usepackage{mathrsfs}
\usepackage{color}
\usepackage{subcaption}

\usepackage{tcolorbox}
\usepackage{tcolorbox}

\usepackage{cancel}

\usepackage[hidelinks]{hyperref}

\newcommand{\Nc}{\ensuremath{N_c}\xspace}
\newcommand{\LONc}{\ensuremath{\text{LO-in-\Nc}}\xspace}
\newcommand{\NLONc}{\ensuremath{\text{NLO-in-\Nc}}\xspace}
\newcommand{\NNLONc}{\ensuremath{\text{N${}^2$LO-in-\Nc}}\xspace}

\newcommand{\NN}{\ensuremath{N\!N}\xspace}

\newcommand{\oneS}{{{}^{1}\!S_0}}
\newcommand{\threeS}{{{}^{3}\!S_1}}
\newcommand{\oneP}{{{}^{1}\!P_1}}

\newcommand{\Pone}{{{}^{3}\!P_1}}
\newcommand{\Pzero}{{{}^{3}\!P_0}}

\newcommand{\calO}{\ensuremath{\mathcal{O}}}

\newcommand{\calI}{\ensuremath{\mathcal{I}}}

\newcommand{\calC}{\ensuremath{\mathcal{C}}}

\newcommand{\calL}{\ensuremath{\mathcal{L}}}

\newcommand{\1}{\mathbbm{1}}
\newcommand{\Num}{\ensuremath{{\mathcal{N}}}}

\newcommand{\pplus}{\vec{p}_+}
\newcommand{\pminus}{\vec{p}_-}
\newcommand{\ppm}{\vec{p}_\pm}
\newcommand{\vsig}{\vec{\sigma}}
\newcommand{\vtau}{\vec{\tau}}

\newcommand{\Czerotrip}{\ensuremath{C_0^{(^3 \! S_1)}}\xspace}
\newcommand{\Czerosing}{\ensuremath{C_0^{(^1 \! S_0)}}\xspace}

\newcommand{\ConeP}{\ensuremath{C^{(^1 \! P_1)}}\xspace}
\newcommand{\CPzero}{\ensuremath{C^{(^3 \! P_0)}}\xspace}
\newcommand{\CPone}{\ensuremath{C^{(^3 \! P_1)}}\xspace}
\newcommand{\CPtwo}{\ensuremath{C^{(^3 \! P_2)}}\xspace}

\newcommand{\threeSoneP}{\ensuremath{C^{({\threeS}-{\oneP})}}\xspace}
\newcommand{\oneSthreePscalar}{\ensuremath{C^{(\oneS-\Pzero)}_{\Delta I=0}}\xspace}
\newcommand{\oneSthreePvector}{\ensuremath{C^{(\oneS-\Pzero)}_{\Delta I=1}}\xspace}
\newcommand{\oneSthreePtensor}{\ensuremath{C^{(\oneS-\Pzero)}_{\Delta I=2}}\xspace}
\newcommand{\threeSthreeP}{\ensuremath{C^{(\threeS-\Pone)}}\xspace}

\newcommand\SD{$S$-$D$\xspace}
\newcommand\SP{$S$-$P$\xspace}
\newcommand\PD{$P$-$D$\xspace}

\newcommand{\asing}{\ensuremath{a^{(^1 \! S_0)}}\xspace}

\newcommand{\atrip}{\ensuremath{a^{(^3 \! S_1)}}\xspace}

\newcommand{\Lone}{\ensuremath{{}^{\nopi}\!L_1}\xspace}
\newcommand{\Ltwo}{\ensuremath{{}^{\nopi}\!L_2}\xspace}

\newcommand{\LoneA}{\ensuremath{L_{1,A}}\xspace}
\newcommand{\LtwoA}{\ensuremath{L_{2,A}}\xspace}

\newcommand{\CMs}{\ensuremath{C^{(M)}_s}\xspace}
\newcommand{\CMv}{\ensuremath{C^{(M)}_v}\xspace}

\newcommand{\gnu}{\ensuremath{g_\nu^{\NN}}\xspace}

\newcommand{\mpi}{\ensuremath{m_\pi}}
\newcommand{\MeV}{\ensuremath{\text{MeV}}}

\newcommand{\nopi}{\ensuremath{\pi\hskip-0.40em /}}
\newcommand{\eftnopi}{EFT$_{\nopi}$\xspace}

\newcommand{\Wigner}{\ensuremath{\text{SU(4)}_\text{Wigner}}\xspace}

\usepackage{siunitx}
\begin{document}

\title{Implications of Large-\Nc QCD for the \NN Interaction}

\author{Thomas R. Richardson}
\affiliation{Institut f\"ur Kernphysik and PRISMA$^+$ Cluster of Excellence, Johannes Gutenberg-Universit\"at, 55128 Mainz, Germany}
\affiliation{Department of Physics, Duke University, Durham, NC 27708, USA}
\author{Matthias R. Schindler}
\affiliation{Department of Physics and Astronomy, University of South Carolina, Columbia, SC 29208, USA}
\author{Roxanne P. Springer}
\affiliation{Department of Physics, Duke University, Durham, NC 27708, USA}

\begin{abstract}

We present a method for ordering two-nucleon interactions based upon their scaling with the number of QCD colors, \Nc, in the limit that \Nc becomes large.
Available data in the two-nucleon sector shows general agreement with this ordering, indicating that the method may be useful in other contexts where data is less readily available.
However, several caveats and potential pitfalls can make the large-\Nc ordering fragile and/or vulnerable to misinterpretation.
We discuss the application of the large-\Nc analysis to two- and three-nucleon interactions, including those originating from weak and beyond-the-standard-model interactions, as well as two-nucleon external currents.
Finally, we discuss some open questions in the field. 
\end{abstract}

\preprint{MITP-22-106}

\maketitle

\section{INTRODUCTION}

It has become increasingly recognized that the large-\Nc limit of QCD \cite{tHooft:1973jz}, where \Nc is the number of colors, is a useful tool for understanding some nuclear physics phenomena.
This makes it potentially helpful for prioritizing future experiments, simulations, and calculations. 
In this review we introduce the basics of the method as it applies to two-nucleon interactions. It is built upon techniques first introduced to study the single-baryon sector.  
For reviews of large-\Nc QCD see, e.g., Refs.~\cite{Witten:1979kh,Coleman:1985rnk,Manohar:1998xv,Jenkins:1998wy,Lebed:1998st}.
We will discuss the application to the two-nucleon sector in the standard model (SM), and then generalize to applications that include more bodies as well as beyond-the-standard-model (BSM) physics.

It is arguably quite surprising that a limit of QCD where the number of colors is taken to be very large has anything to say about our $\Nc=3$ world.  For example, one might expect that corrections to large-\Nc predictions would be on the order of 30\% if an expansion in 1/\Nc is driving the corrections.  
Further, some relationships implied by large-\Nc QCD naively appear to contradict experiment.  
We will see in this review that some of these naive expectations do not hold, and explain the origins of ones that do hold.

The methods discussed in this review apply to terms in a Lagrangian or a nuclear potential, for momenta that are are assumed to be independent of \Nc.
While the interactions may not be observables and are not uniquely defined, they nonetheless reflect the underlying QCD dynamics when used in observables that match available two- and few-nucleon experiments.
The large-\Nc analysis has been applied directly to \NN observables \cite{Cohen:2002qn,Cohen:2011hq,Cohen:2013tya}, but only for momenta proportional to \Nc. 
This momentum regime is not the subject of this review.

We will discuss the impact of the large-\Nc analysis in two contexts: purely phenomenological models and effective field theories (EFTs).
Phenomenological models provide parametrizations of the possible spin, isospin, and momentum structures, with coefficients that are typically fit to scattering data and bound state properties.
While the long-range interactions are most commonly given in terms of single-pion exchange,
a variety of parametrizations are used to describe intermediate- and short-range dynamics: some approaches are purely phenomenological (such as the Argonne v18 potential \cite{Wiringa:1994wb}), in which operator structures are multiplied by specific functions that provide a good fit to data; others are formulated in terms of one-boson-exchange (OBE) contributions (such as the CD-Bonn potential \cite{Machleidt:2000ge}).
The OBE pictures does not imply that physical mesons are exchanged between nucleons and may in fact include the exchange of unphysical bosons; the boson exchange simply serves as a convenient way to generate operator structures. 
The adjustable coefficients in the OBE approach also parametrize multi-meson exchange contributions, e.g., two-pion and $\pi$-$\rho$ exchanges \cite{Machleidt:2000ge}.
For the purpose of the large-\Nc analysis, the details of the potential are irrelevant.
The adjustable parameters capture the strong-interaction dynamics that generate the \NN interactions and the impact of the large-\Nc scaling should be reflected in their relative sizes \cite{Kaplan:1996rk}.  

Over the last few decades, EFTs such as pionless EFT (\eftnopi) and chiral EFT (ChEFT) have been used in an attempt to connect nuclear physics to the symmetries of QCD, see, e.g., Refs.~\cite{Epelbaum:2008ga, Machleidt:2011zz, Hammer:2019poc, Epelbaum:2019kcf} for reviews. 
The cornerstone  of any EFT approach is a separation of low and high energy scales such that the effective Lagrangian can be organized as an expansion in small parameters set by the ratios of these scales.
This organization is referred to as power counting.
Additionally, the effective Lagrangian is constructed to reproduce the symmetries of the underlying theory.
\eftnopi is applicable at energies well below the pion mass and parametrizes two- and few-nucleon interactions in a series of contact terms.
In ChEFT, long- and medium-range interactions are given by single- and multi-pion exchanges, while short-range interactions are represented by \NN contact terms. 
Each term in an EFT Lagrangian is accompanied by a so-called low-energy coefficient (LEC) that encodes the underlying high-energy physics.
In addition, EFTs provide unified frameworks in which to describe two- and few-nucleon interactions, the coupling to external currents, and in the case of ChEFT pionic and pion-nucleon interactions. 
As we will argue below, combining the systematic EFT and large-\Nc expansions is particularly powerful.

The large-\Nc analysis discussed in this review establishes relationships amongst the parameters of a given potential model or the LECs of an EFT.
For the potential models, these relationships are expected to be reflected in the relative sizes of the fitted model parameters.
To assess the validity of the large-\Nc relationships in an EFT, the numerical values of LECs need to be known.
In some cases, such as for operators contributing to \NN scattering, the LECs can be accurately determined from experimental data.
Where this is currently not feasible, lattice QCD in principle provides a method to determine the LECs from the underlying SM.
Therefore, lattice QCD can play an important role in validating large-\Nc expectations.
Lattice QCD calculations can also be performed for $\Nc\ne 3$, but the computational cost increases with \Nc. 
For reviews of $SU(\Nc)$ gauge theories and their lattice formulations, see Refs.~\cite{Lucini:2012gg, Lucini:2013qja}. 
While currently no lattice QCD calculations in the two-nucleon sector with $\Nc>3$ exist, an increasing number of results at the physical $\Nc=3$ continue to become available; for a recent review see, e.g., Ref.~\cite{Davoudi:2020ngi}.
These and future results may provide reliable tests of large-\Nc relationships.

There are certain subtleties that should be kept in mind when comparing large-\Nc expectations to data; these will be discussed in detail. 
Some of them are related to the fact that the large-\Nc analysis applies to parameters/LECs, which are not observables.
Further, scales unrelated to large-\Nc dynamics can be relevant in nuclear physics. 
Given the expansion parameter $1/\Nc = 1/3$, these scales may impact how well large-\Nc relationships are reflected in data.
The results of large-\Nc analyses should therefore not be viewed as precise predictions, but rather as indicating trends.
Nonetheless, the large-\Nc expansion provides an important and useful tool for illuminating the connection between two- and few-nucleon interactions and the underlying theory of QCD.

\section{ELEMENTS OF THE LARGE-\Nc EXPANSION}
\label{sec:aspects}

In this section we provide the tools and describe the steps needed to determine how the large-\Nc limit of QCD  can be used to establish a hierarchy among two-nucleon interactions.
These interactions may involve SM and/or BSM physics.  
We will ultimately also discuss processes that may  involve  external currents and/or pions in addition to nucleons.

\subsection{Determining large-\Nc scaling}
Because of the nonperturbative nature of QCD, it is challenging to calculate matrix elements of quark-level operators between baryon states. But it is possible to determine how such matrix elements scale with the number of colors \Nc. To determine how this impacts nuclear physics, the strategy is to map the \Nc scaling of the QCD matrix elements onto an effective field theory (EFT) expressed in terms of baryon  fields or, equivalently, onto a potential.  The low energy coefficients (LECs) in that EFT or potential inherit the scaling of the QCD matrix elements.

This mapping process relies on some key results from the single-baryon sector. 
First, baryons have a wave function that is totally antisymmetric with respect to interchange of its constituent fermions. 
Since the quarks in the baryon must form a color singlet, the baryons must be composed of \Nc quarks, resulting in a baryon mass that grows as \Nc \cite{Witten:1979kh}.
Second, Witten showed that the baryon-meson scattering amplitude at fixed meson energy is $O(1)$,
and that additional insertions of mesons at a single vertex are suppressed; baryon-meson scattering is dominated by a single meson at each vertex \cite{Witten:1979kh}.
Crucially, the baryon-meson interaction vertex is $O(\sqrt{\Nc})$, which would lead to an amplitude that scales as \Nc. 
To obtain an amplitude that is $O(1)$, an infinite tower of degenerate baryon states is required in the large-\Nc limit \cite{Gervais:1983wq, Gervais:1984rc,Dashen:1993ac, Dashen:1993as}.
This tower of states furnishes a representation of a \emph{contracted} SU(4)  symmetry with the Lie algebra given by \cite{Dashen:1993ac, Dashen:1993as, Dashen:1993jt} 
    \begin{align}\label{eq:ContractedCommRel}
    [S^i,X^{ja}] & = i \epsilon^{ijk} X^{ka}\, , & [I^a,X^{ib}] & = i \epsilon^{abc} X^{ic}\, , & [X^{ia},X^{jb}] & = 0 \, ,
    \end{align}
where $S^i$ and $I^a$ are the generators of spin and isospin, respectively, for which the usual commutation relations hold.
The contracted group is related to the full SU(4) algebra 
    \begin{align}
    [S^i,S^j] & = i \epsilon^{ijk} S^k\, , & [I^a,I^b] & = i \epsilon^{abc} I^c\, , & [S^i,I^a] & = 0\, ,\\
    [S^i,G^{ja}] & = i \epsilon^{ijk} G^{ka}\, , & [I^a,G^{ib}] & = i \epsilon^{abc} G^{ic}\, , & [G^{ia},G^{jb}] & = \frac{i}{4} \delta^{ij} \epsilon^{abc} I^c + \frac{i}{4} \delta^{ab} \epsilon^{ijk} S^k\, , \label{eq:GAlgebra}
    \end{align}
through the contraction 
    \begin{equation}
        X^{ia} = \lim_{\Nc \to \infty} \frac{G^{ia}}{\Nc} \, .
    \end{equation}
One representation for the operators $\{S,I, G\}$ \cite{Dashen:1993as,Dashen:1993jt,Dashen:1994qi,Carone:1993dz} uses  bosonic-quark operators,
\begin{align}
    S^i = q^\dagger \frac{\sigma^i}{2}q,\quad I^a = q^\dagger \frac{\tau^a}{2}q,\quad G^{ia}  = q^\dagger \frac{\sigma^i\tau^a}{4}q \, ,
\end{align}
where the $q^\dagger$ and $q$ are creation and annihilation operators, respectively, for the light quark flavors and $\sigma^i$ ($\tau^a$) are Pauli matrices in spin (isospin) space.
These bosonic quarks are not QCD quarks.
For baryons, the antisymmetry in the \Nc quark colors allows these quarks to be colorless and bosonic \cite{Dashen:1994qi}.
The large-\Nc scaling of the matrix elements of the operators $\{S,I, G\}$ contributes to the large-\Nc scaling of baryon sector quantities.
For physical baryons with spin and isospin of $O(1)$, such as the nucleons considered in the following, matrix elements of the $S$ and $I$ operators are $O(1)$, but the matrix elements of the $G$ operators are $O(\Nc)$. 
Also, since baryons are composed of \Nc quarks, the matrix elements of the quark number operator $\Num = q^\dagger q$ scale as \Nc.
In summary, the following rules hold:
\begin{tcolorbox}
Isospin/spin scaling rules: \begin{align}
\label{eq:SIGScale}
    \langle N' | \frac{ G^{ia}}{\Nc} | N \rangle & \sim 1\, , 
    & \langle N | \frac{\Num}{\Nc} | N \rangle & \sim 1 \, , 
    &\langle N' | \frac{ S^{i}}{\Nc} | N \rangle & \sim \Nc^{-1}\, , 
    & \langle N' | \frac{ I^a}{\Nc} | N \rangle & \sim \Nc^{-1} \ ,
\end{align} 
\end{tcolorbox}
\noindent where the $|N\rangle$ are the nucleon states.

In the two-nucleon sector (a parallel argument holds for three and more nucleons) the objective is to compare a two-nucleon matrix element of an operator $\tilde\calO$ involving the quark and gluon fields of QCD to a two-nucleon matrix element of an operator $\calO_\text{EFT}$ in the EFT/potential that has the same symmetry properties and quantum numbers. 
The large-\Nc scaling of the quark-level matrix element is then mapped onto the LEC $C(\Nc)$ that comes with $\calO_\text{EFT}$,
\begin{equation}
\label{matching}
        \langle N_\gamma N_\delta | \tilde \calO | N_\alpha N_\beta  \rangle \rightarrow C(\Nc) \langle N_\gamma N_\delta | \calO_{\text{EFT}} | N_\alpha N_\beta\rangle \, ,
\end{equation}
\noindent where $\alpha, \beta, \gamma,$ and $\delta$ hold the spin-isospin indices of the external nucleon fields. 
The operator $\tilde\calO$ contains the quark and gluon fields of QCD, but can also encompass other SM and BSM degrees of freedom.
$\tilde \calO$ can be characterized by the number of quarks it acts upon; an $m$-body operator $\tilde\calO^{(m)}$ acts on $m$ quarks. 
In the baryon sector, these operators can be expanded in terms of the operators $\{S,I,G \}$ \cite{Dashen:1994qi,Carone:1993dz,Luty:1993fu},
\begin{equation}
\label{eq:spin-flavor}
    \tilde\calO^{(m)} =  \Nc^m \sum_{n=0}^{\Nc} \sum_{s+t \leq n} c_{stn} \left( \frac{S}{\Nc} \right)^s \left( \frac{I}{\Nc} \right)^t \left( \frac{G}{\Nc} \right)^{n-s-t} \, .
\end{equation}
The coefficients $c_{stn}$ depend only on the dynamics of QCD, independent of the \Nc scaling arising from the SU(4)-symmetry, and can carry spatial indices arising from momenta.
In particular, $m=1$ operators in the QCD Hamiltonian have a large-\Nc expansion in the baryon sector of the form 
\begin{align}
\label{eq:Hartree}
    H = \Nc \sum_{n=0}^{\Nc} \sum_{s+t \leq n} v_{stn} \left( \frac{S}{\Nc} \right)^s \left( \frac{I}{\Nc} \right)^t \left( \frac{G}{\Nc} \right)^{n-s-t} \, ,
\end{align}
where the coefficients $v_{stn}$ are functions of momenta and play the role of the $c_{stn}$ coefficients in Eq.~\eqref{eq:spin-flavor}.
On the right-hand side (RHS) of Eqs.~\eqref{eq:spin-flavor} and \eqref{eq:Hartree}, the suppressed indices of the operators and the coefficients are contracted in order to reproduce the symmetry properties of the operator on the left-hand side (LHS). 
Many of the operators that occur in the expansion on the RHS side can be eliminated using operator reduction rules \cite{Dashen:1994qi}, greatly simplifying the form of the expansion.
For example, products of $G^{ia}$ acting on the same nucleon in which any indices are contracted with the Kronecker delta and/or the Levi-Civita tensor can be reduced to terms containing fewer factors of $G$ \cite{Kaplan:1996rk}.
In the two-nucleon sector, terms with multiple factors of the form $\left(G_1^{ia} G_2^{ia}\right)$, where the subscript denotes the nucleon the operator acts on, can be reduced to a form that at leading order in \Nc (\LONc) contains only a single factor of the identity operator and of $G$ \cite{Phillips:2013rsa}.

In the large-\Nc limit, the dominant contributions to a two-nucleon matrix element are those that factorize into the products of two single-nucleon matrix elements of terms in the expansion of Eq.~\eqref{eq:spin-flavor}. For a discussion of this point, see, e.g., Ref.~\cite{Kaplan:1995yg}.
\begin{tcolorbox}
Factorization rule for two-baryon matrix elements
\begin{align}\label{eq:factorization}
        \langle N_\gamma(p_1') N_\delta (p_2')| \tilde\calO | N_\alpha(p_1) N_\beta(p_2)\rangle  \to  
        \langle N_\gamma(p_1') | \calO_\text{I} | N_\alpha(p_1) \rangle
        \langle N_\delta(p_2')| \calO_\text{II} | N_\beta(p_2)\rangle + \text{crossed}\, .
    \end{align}
\end{tcolorbox}
The large-\Nc scaling that comes from spin-isospin properties on the RHS is determined from the single-nucleon matrix elements of Eq.~\eqref{eq:SIGScale}.
In practice, this means that the large-\Nc scaling of a two-nucleon  matrix element inherits the scaling of the single-nucleon matrix elements of the operators $\calO_\text{I}$ and $\calO_\text{II}$, and an overall factor of \Nc is removed to account for the scaling of the Hamiltonian in Eq.~\eqref{eq:Hartree}.
\begin{tcolorbox}
Remove a factor of \Nc to account for the overall scaling with \Nc in Eq.~\eqref{eq:Hartree}.
\end{tcolorbox}

The nonrelativistic momenta in the coefficients $v_{stn}$ are assumed to be independent of \Nc: $p \sim O(\Nc^0)$ \cite{Kaplan:1996rk}.
An additional source of large-\Nc suppression for two-nucleon matrix elements comes from momenta that arise from relativistic corrections. 
To determine the impact on the mapping in Eq.~\eqref{matching}, we use the center-of-mass (COM) frame and introduce the combinations  
\begin{align}
    \ppm \equiv \vec{p}^{\,\prime}\pm \vec{p} \,,
\end{align}
where 
\begin{align}
    \vec{p}^{\,\prime} = \vec{p}^{\,\prime}_1 - \vec{p}^{\,\prime}_2, \quad 
    \vec{p} = \vec{p}_1-\vec{p}_2 \, 
\end{align}
are the momenta labeled in Eq.~\eqref{eq:factorization}. 
For \NN interactions in the t channel, the momentum structure $\pplus$ originates from a relativistic correction and is accompanied by a factor of $1/M \sim 1/\Nc$ \cite{Kaplan:1996rk}.
Similarly, in the u channel the momentum $\pminus \sim 1/\Nc$. 
Results obtained from the t and u channels are equivalent.
In the following we only consider the t channel, leading to
\begin{tcolorbox}
Momentum scaling rule for two-baryon matrix elements
\begin{align}
\label{eq:MomScale}
    \pminus \sim 1 \, , \quad \pplus \sim \Nc^{-1} \, .
\end{align}
\end{tcolorbox}
\noindent This rule holds for an analysis in the two-nucleon COM frame. 
For momenta in the three-nucleon system see Sec.~\ref{sec:three}.

An additional source of suppression in the scaling of the LECs occurs if pions are present.
Each pion at a vertex is accompanied by a factor $1/F_\pi \sim 1/\sqrt{\Nc}$, where $F_\pi$ is the pion decay constant.
\begin{tcolorbox}
Pion suppression rule
\begin{center} Each $n$-pion vertex leads to a suppression by $\Nc^{-n/2}$.
\end{center}
\end{tcolorbox}

There is a further subtlety related to Fierz transformations.
These transformations can be used to eliminate redundant terms and reduce a Lagrangian to a minimal form. This minimal form is not unique;  the choice of operators to retain is based on convenience.  For a large-\Nc analysis, the most convenient choice is to retain those operators that are largest in \Nc.
If that choice is not made, applying the above rules to an EFT or potential that has been reduced by Fierz transformations can obscure the correct \Nc scaling of the LECs \cite{Schindler:2015nga,GirlandaCD15}.
For example, an LEC of the minimal basis might appear to be suppressed, but it may have been obtained by eliminating an operator with an LEC that is lower order in the large-\Nc expansion. 
To avoid errors in determining the \Nc scaling of operators, the most straightforward procedure is to begin with the most general, over-complete set of operators that is consistent with the desired symmetries, power counting, etc. 
Then the large-\Nc scaling of each operator in the over-complete set is determined using the ingredients above.
As a final step, Fierz transformations can be performed to reduce the number of operators while retaining those that are dominant in \Nc scaling.
\begin{tcolorbox}
    Perform the large-\Nc analysis prior to the application of Fierz transformations.
\end{tcolorbox}

\subsection{Complications, caveats, and cautions}
When using the approach outlined above it is important to be aware of the following:
\begin{itemize}
\item The $\Delta$ and nucleons are degenerate in the large-\Nc limit and the presence of the $\Delta$ is required in the single-baryon sector in the derivation of the contracted SU(4) algebra \cite{Dashen:1993ac, Dashen:1993as, Gervais:1983wq, Gervais:1984rc,Dashen:1993jt}. 
A consistent large-\Nc treatment of nuclear physics would seem to require the explicit inclusion of $\Delta$ particles in intermediate states. 
In practice this does not seem to be necessary, meaning that so far the counting rules applied without a dynamical $\Delta$ do not disagree with experiment. We discuss this issue more thoroughly in Ref.~\cite{delta:inprep}, also see the discussions in Refs.~\cite{Kaplan:1996rk,Beane:2002ab}. 
\item The large-\Nc rules above  only provide upper bounds on the \Nc scalings. It is possible that additional cancellations (possibly due to an unidentified symmetry, whether driven by QCD dynamics or otherwise) reduce this scaling.
Additionally,  enhancements that are independent of \Nc could obscure the large-\Nc counting, if, for example, accidental large coefficients appear in the expansion. 
Any of these could impact how closely physical quantities match the predicted SU(4) large-\Nc hierarchy in a world with $\Nc = 3$. 

\item It was argued by Witten \cite{Witten:1979kh} that the baryon-baryon interaction should be $O(\Nc)$ and that the baryon-baryon scattering amplitude has a smooth (and nontrivial) limit for momenta $p\sim O(\Nc)$.
For $p\sim O(\Nc^0)$ as assumed here, the nucleon kinetic energy $p^2/2M$ is of order $O(\Nc^{-1})$ and therefore much smaller than the potential energy.
There does not exist a smooth large-\Nc limit for the scattering amplitude.
Reference \cite{Kaplan:1996rk} argues that, nonetheless, hierarchies derived on the basis of the contracted SU(4) symmetry in the baryon sector with $p\sim O(\Nc^0)$  are expected to hold.

\item The procedure described above determines the large-\Nc scaling of the LECs, which are not themselves physical observables.
In particular, as is typical in an EFT, the LECs are often renormalization-scheme and -point dependent. 
Different LECs evolve differently with the renormalization scale, and a large-\Nc hierarchy among them might only be manifest for certain domains of the renormalization point. See, for example, the discussion in Sec.~\ref{sec:sym_preserve}.

\item Because LECs are not observables, it is rarely consistent to take LECs from one theory and use them in another. 
For example, the renormalization scale dependence in one theory may not be the same as in another.
Therefore, the \Nc scaling of LECs in one theory cannot reliably be used to predict the \Nc scaling of LECs in a different theory. 
Instead, each theory should be independently analyzed.

\item In the large-\Nc limit, the $\Delta$-nucleon mass difference tends to zero, while the pion mass tends to zero in the chiral limit.
In ChEFT, some observables depend on the ratio of these two quantities. Therefore, when applying large-\Nc counting to ChEFT it is important to recognize that the chiral and large-\Nc limits do not commute, see, e.g., \cite{Jenkins:1991ne,Cohen:1996zz}. 

\end{itemize}
Despite these caveats, the constraints that arise from applications of the large-\Nc rules to the \NN interactions  have so far been found to be in general agreement with data as long as appropriate values of $\mu$ are used.

\subsection{Summary}
The process for determining the large-\Nc scaling of LECs in an EFT or potential is:
    \begin{tcolorbox}
    \begin{enumerate}
        \item Derive the most general and over-complete set of operators consistent with the symmetries of the system; in particular, no Fierz reductions should be applied at this stage.
        \item Use the factorization property in conjunction with the \Nc scaling of single-baryon matrix elements to determine how the spin-isospin structure of each operator contributes to the \Nc scaling.
        \item Remove a factor of \Nc to account for the overall scaling of the operator in Eq.~\eqref{eq:Hartree}.
        \item Determine suppressions due to any relativistic corrections from momenta.
        \item Include suppressions due to any pion interactions.
        \item  Optionally, eliminate higher-order-in-\Nc redundant terms through Fierz transformations. 
    \end{enumerate}
    \end{tcolorbox}

\section{SYMMETRY-PRESERVING NUCLEON-NUCLEON INTERACTIONS}\label{sec:sym_preserve}

The rules of Sec.~\ref{sec:aspects} will now be applied to strong interactions. In the subsections below we discuss leading and higher orders in both \eftnopi and large-\Nc counting, present the implications of the SU(4) spin-flavor symmetry, compare the large-\Nc results to experimental and lattice data, and then briefly summarize some related results concerning approximate symmetries in the two-nucleon interaction and the deuteron binding energy.

\subsection{Aspects of combining the large-\Nc and \eftnopi expansions}

\subsubsection{Leading order in \eftnopi}

The simplest example of using  large-\Nc scaling techniques to prioritize operators is to consider the leading order strong interactions in \eftnopi.
The most general Lagrangian of zero-derivative two-nucleon contact operators that are invariant under rotations and isospin transformations contains four possible operators,  \begin{equation}\label{eq:four.strong}
        \calL = - \frac{1}{2} C_{\1 \cdot \1} \left( N^\dagger N \right)^2 - \frac{1}{2} C_{\sigma \cdot \sigma} \left( N^\dagger \vec \sigma N \right)^2 - \frac{1}{2} C_{\tau \cdot \tau} \left( N^\dagger \vec \tau N \right)^2 - \frac{1}{2} C_{G \cdot G} \left( N^\dagger \sigma^i \tau^a N \right)^2 \, ,
    \end{equation}
where the LEC subscripts echo the operator structure they multiply.
Mapping the scalings of the large-\Nc matrix elements in Eq.~\eqref{eq:SIGScale} onto the LECs of the corresponding operators with the same symmetry properties and removing an overall factor of \Nc 
yields
\begin{equation}
\label{eq:overcomplete_scaling_pionless}
        C_{\1 \cdot \1}, \ C_{G \cdot G} \sim O(\Nc)\, , \quad \quad C_{\sigma \cdot \sigma}, \ C_{\tau \cdot \tau} \sim O(1/\Nc) \, .
    \end{equation}
Two of the four operators are redundant and can be removed by Fierz transformations.
The resulting Lagrangian can be chosen to be
\begin{align}
\label{eq:LOLag}
    \calL = - \frac{1}{2} C_S \left( N^\dagger N \right)^2 - \frac{1}{2} C_T \left( N^\dagger \vec \sigma N \right)^2 \, ,
\end{align}
with 
\begin{align}
        C_S =  C_{\1 \cdot \1} - 3 C_{G \cdot G} - 2 C_{\tau \cdot \tau} \, ,\quad \quad
        C_T =  C_{\sigma \cdot \sigma} - C_{\tau \cdot \tau} \label{eq:C_scaling}\, .
    \end{align}
The scalings of Eq.~\eqref{eq:overcomplete_scaling_pionless} dictate that
\begin{align}
\label{eq:CST-scaling}
    C_S \sim O(\Nc) \, ,\quad  C_T \sim O(1/\Nc)\, .
\end{align}
While this example exhibits some of the general features of a large-\Nc analysis in the \NN sector, some potential subtleties discussed in Sec.~\ref{sec:aspects} are absent. 
For instance, the Fierz transformations for this example do not obscure the large-\Nc scaling, due to the simplicity of the form of the operators.
That is, the application of the large-\Nc counting rules directly to the minimal Lagrangian of Eq.~\eqref{eq:LOLag} results in the correct large-\Nc scalings of Eq.~\eqref{eq:CST-scaling}.
Further, the LO Lagrangian does not contain any factors of momentum that can impact the large-\Nc scaling and, being a pionless theory, no factors of the pion decay constant enter.
In the next sections, we present several examples where the situation is more complex; Fierz transformations have a nontrivial impact on apparent \Nc scaling and momentum factors and pions are present.

\subsubsection{SU(4)-symmetric Lagrangian}
Before considering further applications of the general approach, we show that the vanishing of $C_T$ in the large-\Nc limit also follows more directly from the SU(4) symmetry that emerges in the baryon sector \cite{Kaplan:1995yg}. 
In that limit, the nucleon and $\Delta$ states form the 20-dimensional fundamental representation $\Psi^{ABC}$, where the capital Latin superscripts are SU(4) indices. The lowest-dimensional Lagrangian describing the interactions between two baryons is given by \cite{Kaplan:1995yg}
    \begin{align}\label{eq:SU4Lag}
        \calL_{6} = -\frac{1}{f_\pi^2} \left[ a (\Psi^\dagger_{ABC}\Psi^{ABC})^2 
        + b \Psi^\dagger_{ABC}\Psi^{ABD} \Psi^\dagger_{EFD}\Psi^{EFC} \right]\, .
    \end{align}
Expanding the baryon fields $\Psi$ in terms of nucleons and $\Delta$'s and matching to the Lagrangian of Eq.~\eqref{eq:LOLag} gives
\begin{align}
\label{eq:CST-ab}
    C_S = \frac{2}{f_\pi^2}\left( a-\frac{b}{27} \right), \quad C_T = 0 \, ,
\end{align}
which is consistent with the scaling of Eq.~\eqref{eq:CST-scaling} as $\Nc \to \infty$.

\subsubsection{Comparison with data}
To test the large-\Nc expectation that $C_T=0$, it is convenient to consider the \NN interactions in a partial-wave basis. The Lagrangian in Eq.~\eqref{eq:LOLag} is equivalent to LO \NN interactions in the $\oneS$ and $\threeS$ channels, with corresponding LECs
\begin{align} \label{eq:C0TwoBases}
    \Czerosing = (C_S - 3C_T),
    \quad \Czerotrip = (C_S + C_T) \, ,
\end{align}
or $\Czerosing = \Czerotrip$ at LO in the large-\Nc expansion.
However, the LECs are not observables; while $\Czerosing$ and $\Czerotrip$ can be related to the spin-singlet and spin-triplet scattering lengths $\asing$ and $\atrip$, they are also dependent on a renormalization scheme and point.
In the PDS scheme \cite{Kaplan:1998tg}, the relationship between the S-wave LEC and the scattering length in channel $s$ is
\begin{align}
    C_0^{(s)} = \frac{4\pi}{M} \frac{1}{\frac{1}{a^{(s)}} - \mu}\, ,
\end{align}
with $\mu$ the renormalization parameter.
As pointed out in Ref.~\cite{Kaplan:1995yg}, choosing a renormalization point such that the LECs are dominated by the values of the scattering lengths hides the expected large-\Nc scaling, because the physical values of $a^{(s)}$ are fine-tuned and anomalously large,
\begin{align}
    \asing \approx -23.7 \text{ fm}, \quad \atrip \approx 5.41 \text{ fm} .
\end{align}
Using $\mu = 0$, the ratio $\Czerosing/\Czerotrip \approx - 4.4$.  
Not only is the magnitude far from unity, but even the sign disagrees with the large-\Nc result.
However, setting $\mu$ equal to the pion mass 
gives
\begin{align}\label{eq:C0ratio}
    \left.\frac{\Czerosing(\mu = \mpi)}{\Czerotrip(\mu = \mpi)}\right|_\LONc \approx 0.7.
\end{align}
The 30\% deviation from unity is consistent with the large-\Nc expectation. 
While corrections to the \LONc expressions for $C_S$ and $C_T$ are suppressed by $1/\Nc^2$, the factor of 3 in Eq.~\eqref{eq:C0TwoBases} results in corrections to Eq.~\eqref{eq:C0ratio} that are approximately $4/\Nc^2 \sim 1/\Nc$ for $\Nc=3$. 
In the large-\Nc basis, the corresponding ratio 
\begin{align}
    \frac{C_T}{C_S} \approx 0.08
\end{align}
is consistent with the expected $1/\Nc^2$ suppression \cite{Schindler:2018irz}.
Reference \cite{Kaplan:1995yg} uses a square-well model for the \NN potential to find $C_T/C_S \approx 0.05$.

\subsubsection{Comparison with lattice QCD}

In addition to experiment, large-\Nc expectations can also be compared to lattice QCD results.
Reference \cite{NPLQCD:2013bqy} determines the scattering lengths and effective ranges in the two $S$-wave channels for SU(3)-symmetric quark masses at the physical value of the $s$ quark mass, corresponding to  a pion mass $m_\pi \approx 806 \,\MeV$. 
The results are consistent with the vanishing of $C_T$ as expected in the large-\Nc limit. 
In fact, at this unphysical pion mass, the $S$-wave scattering lengths found in Ref.~\cite{NPLQCD:2013bqy} have the same sign and are close in magnitude, and the equality of the interactions in the two S-wave channels does not depend on choosing a suitable renormalization point for the LECs.
Similar results are also reported in Ref.~\cite{Horz:2020zvv}, with the two scattering lengths of equal sign and similar magnitude at a slightly lighter pion mass.
The suppression of $C_T$ relative to $C_S$ is also observed in the lattice QCD results in Ref.~\cite{Detmold:2021oro}.\footnote{Note the different definition of $C_S$ and $C_T$ in Ref.~\cite{Detmold:2021oro}, which correspond to \Czerosing and \Czerotrip in the conventions used here.} 

Generalizing from SU(2) to SU(3) to describe the three-flavor spin-1/2 baryon-baryon interactions provides a particularly interesting example of the interplay of large-\Nc and lattice QCD methods. 
For three flavors there are six independent contact terms at LO in the EFT power counting, with the  Savage-Wise coefficients \cite{Savage:1995kv} as the corresponding LECs.
The LO Lagrangian in the large-\Nc expansion still has the form of Eq.~\eqref{eq:SU4Lag}, but with the fields now representing spin-1/2 octet and spin-3/2 decuplet fields \cite{Kaplan:1995yg}; there are still two independent parameters, only now they constrain the six Savage-Wise LECs, resulting in a baryon-level SU(6) spin-flavor symmetry. 
An analogous SU(4) spin-isospin symmetry, referred to as Wigner symmetry, arises in the two-flavor case, see Sec.~\ref{sec:approxsym} for additional discussion.
Further, only one Savage-Wise LEC depends on $a$, while the contributions proportional to $b$ are numerically suppressed. 
If $b\lesssim a$, the terms proportional to $b$ can be neglected at \LONc, leading to an accidental SU(16) symmetry in the baryon-baryon interactions \cite{Kaplan:1995yg}. 
The lattice QCD analysis of these interactions was performed in Ref.~\cite{Wagman:2017tmp}, again using SU(3)-symmetric quark masses corresponding to $m_\pi \approx \SI{806}{MeV}$. 
The results for the extracted scattering parameters are consistent not only with the approximate SU(6) spin-flavor symmetry related to the large-\Nc limit, but also with the larger accidental SU(16) symmetry. 
These interactions were also analyzed for lighter $u$ and $d$ quark masses, corresponding to $m_\pi \approx 450 \,\MeV$ and $m_K \approx 596  \,\MeV$ \cite{NPLQCD:2020lxg}. 
The interactions again exhibit an SU(6) spin-flavor symmetry for these quark masses. 
They are also consistent with the accidental SU(16) symmetry, but this feature is less pronounced than at the heavier pion mass.
This apparent SU(16) symmetry has been related to the suppression of entanglement in low-energy baryon-baryon scattering \cite{Beane:2018oxh}.

\subsubsection{Higher order terms in \eftnopi} \label{sec:twoderivative}

The \NN interaction terms at the next order in \eftnopi illustrate two additional aspects of the large-\Nc analysis: the terms contain two factors of momenta, requiring the use of the rules of Eq.~\eqref{eq:MomScale}; and na\"ive application of the large-\Nc scaling rules to a minimal form of the Lagrangian would result in incorrect \Nc scaling assignments. 
For example, two \LONc operators are proportional to $\vec\tau_1 \cdot \vec\tau_2$.
If they are eliminated through Fierz transformations in favor of terms that only contain isospin identity operators, their contributions are subsumed in LECs that are seemingly of higher order in the 1/\Nc expansion. 

The list of operators with two derivatives at \LONc and \NLONc is given in Ref.~\cite{Schindler:2018irz}.
There are three \LONc operators, all proportional to two factors of $\pminus$.
The eight \NLONc terms  are suppressed by $1/\Nc^2$. Not all of these terms are independent; four can be removed by Fierz transformations. The suppression by $1/\Nc^2$ is in agreement with the general observation of Ref.~\cite{Kaplan:1996rk} that for the symmetry-preserving \NN interaction the expansion is in powers of $1/\Nc^2$ instead of 1/\Nc.

In the partial-wave basis there are seven operators with two derivatives, corresponding to higher-order (in the momentum expansion) corrections to the two $S$-wave interactions, an \SD mixing term, and four $P$-wave interactions.
The \Nc counting is performed in the basis where operators have the form $(N^\dagger O_1 N)(N^\dagger O_2 N)$, such as those in Eq.~\eqref{eq:four.strong}. We call this the ``large-\Nc basis.''
In this basis, there are only three independent \LONc operators.  
All seven partial-wave LECs overlap with operators that are \LONc, leading to relationships amongst them.
These were analyzed in Ref.~\cite{Schindler:2018irz}.
To check how well the analysis works we compare to experimental phase shifts (to the extent possible given the $\mu$ dependence of some LECs)  and check whether LECs occurring within a given order in the large-\Nc expansion are of similar size.

The values of the $P$-wave LECs ($\mu$-independent at this order in the \eftnopi power counting) extracted from low-energy scattering phase shifts are \cite{Schindler:2018irz}
\begin{equation}\label{eq:Pwave_exp}
  \CPzero = 6.6\, \text{fm}^4 \ , \quad \CPone = -6.0\, \text{fm}^4 \ , \quad
  \CPtwo = 0.57\, \text{fm}^4 \ , \quad \ConeP  = -22\, \text{fm}^4 \ .
\end{equation}
The large differences in these values would seem to contradict the result that all $P$-wave LECs receive \LONc contributions. 
However, the large-\Nc analysis is performed on the operators in the large-\Nc basis. 
The transition to the partial-wave basis  of the LECs $C^{(^{2S+1} \! P_J)}$ introduces relative factors as large as 36, matching the relative suppression seen in Eq.~\eqref{eq:Pwave_exp}.
The three leading-in-\Nc terms in the large-\Nc basis, when fit to the $P$-waves, yield LECs of similar relative size.
The same trend regarding the relative sizes of LECs in different bases is also seen in the magnetic and axial two-nucleon currents \cite{Richardson:2020iqi}; see Sec.~\ref{sec:magnetic}.

Because the LECs associated with $S$-waves are $\mu$ dependent (and hence not physical observables), we must choose a $\mu$ value to compare all seven of the two-derivative LECs  with experiment.
Reference \cite{Schindler:2018irz} found that the large-\Nc  relationships are reasonably consistent with experiment only  for a limited domain of the renormalization point $\mu$ (in the PDS scheme), analogous with the findings at LO.
A fit of the seven LECs at $\mu = 120 \, \text{MeV}$ to scattering data does not show a suppression of the \NNLONc terms relative to the \LONc terms. 
This was traced back to the unnaturally small value of the \SD-mixing seen in experiment, which is nominally of \LONc.
Artificially increasing the value of the \SD LEC by a factor of 3 improves the agreement in the pattern of the fitted values with the large-\Nc hierarchy. 
This example shows that the large-\Nc expansion does not necessarily capture all relevant physics and that other mechanisms can impact the relative sizes of LECs.

\subsection{General parameterization of the \NN potential}

Using the scaling relations of nucleon matrix elements in Eq.~\eqref{eq:SIGScale}, Ref.~\cite{Kaplan:1996rk} considered the large-\Nc analysis of a general parameterization of the \NN potential,
\begin{align}
\label{eq:V_general}
    V_{\NN} & = V^0_0 + V^0_\sigma \vec{\sigma}_1 \cdot \vec{\sigma}_2 + V^0_\text{LS} \vec{L} \cdot \vec{S} + V^0_T S_{12} + V^0_Q Q_{12}
    + \left(V^1_0 + V^1_\sigma \vec{\sigma}_1 \cdot \vec{\sigma}_2 + V^1_\text{LS} \vec{L} \cdot \vec{S} + V^1_T S_{12} + V^1_Q Q_{12} \right) \vec{\tau}_1\cdot\vec{\tau}_2\, ,
\end{align}
where $\vec{L}$ is the total orbital angular momentum, $\vec{S}$ the total spin of the two-nucleon state, and
\begin{align}
    S_{12} & \equiv 3\vec{\sigma}_1 \cdot \hat{r}_1 \vec{\sigma}_2 \cdot \hat{r}_2 - \vec{\sigma}_1 \cdot\vec{\sigma}_2 \, , \\
    Q_{12} & \equiv \frac{1}{2} \left\{ \vec{\sigma}_1 \cdot \vec{L}, \vec{\sigma}_2 \cdot \vec{L} \right\} \, .
\end{align}
Reference~\cite{Kaplan:1996rk} found the large-\Nc scalings of the functions $V^i_j$ to be
\begin{align}
    V^0_0 \sim V^1_\sigma \sim V^1_T & \sim \Nc \, ,\\
    V^1_0 \sim V^0_\sigma \sim V^0_\text{LS} \sim V^1_\text{LS} \sim V^0_T \sim V^1_Q &\sim 1/\Nc \, \\
    V^0_Q & \sim 1/\Nc^3 \, ,
\end{align}
illustrating that the expansion of the strong interaction potential is in powers of $1/\Nc^2$ instead of $1/\Nc$. This makes the large-\Nc expansion more useful than na\"ively expected.

\subsubsection{Comparison to phenomenological potentials}
The large-\Nc expectations can be tested by comparison with phenomenological models when these are considered as parametrizations of \NN data. 
As pointed out in Ref.~\cite{Kaplan:1996rk}, the success of this comparison does not depend on the particular details of a model. 
For each parametrization, after taking into account the large-\Nc scaling from momenta and possible factors of the nucleon mass, the large-\Nc hierarchy should be reflected in the relative sizes of the parameters that are fit to data.
The authors of \cite{Kaplan:1996rk} compared the derived large-\Nc scalings with the relative sizes of parameters fit to data in the Nijmegen \NN potential \cite{Nagels:1977ze,Stoks:1994wp}.
In general, there is good agreement between the anticipated large-\Nc ordering and the Nijmegen phenomenological parameters.\footnote{See Figure 3 in the arXiv version of Ref.~\cite{Kaplan:1996rk}; the $y$ axis of the plot appears to be slightly offset in the published version.}
Two parameters from the Nijmegen potential are smaller than expected from the large-\Nc analysis.
However, as discussed in Sec.~\ref{sec:aspects}, the large-\Nc scalings only provide upper bounds.
In addition to the Nijmegen potential, Ref.~\cite{Riska:2002vn} extended the comparison with phenomenological potentials to also include the Argonne v18 \cite{Wiringa:1994wb},  CD-Bonn \cite{Machleidt:2000ge}, and Paris \cite{Lacombe:1980dr} potentials. 
The analysis showed that the strength of the various components of these different potentials in general match the large-\Nc expectations, supporting the claim that the success of the comparison does not depend on the choice of the potential model.

\subsubsection{The meson exchange picture and the large-\Nc expansion}\label{sec:mesonexchange}
A common parametrization of the \NN potential is in terms of meson exchanges.
This is useful for parameterizing experiment, but only pion exchanges are physical at these energies. 
The relationship between the meson-exchange picture and  the large-\Nc expansion was studied in Refs.~\cite{Banerjee:2001js,Belitsky:2002ni,Cohen:2002im}.
For a two-nucleon interaction given in terms of meson exchanges, the known scalings of meson-nucleon vertices can be combined to derive the expected large-\Nc scaling of the potential.
Reference \cite{Banerjee:2001js} shows that the large-\Nc scalings extracted in Ref.~\cite{Kaplan:1996rk} are consistent with a potential based on meson exchanges up to two-meson-exchange diagrams.
This consistency relies on cancellations between different diagrams and requires the inclusion of $\Delta$ intermediate states, which are not considered in Ref.~\cite{Kaplan:1996rk}.
The consistency between the meson-exchange and the large-\Nc pictures seems to break down when considering the exchanges of three and more mesons \cite{Belitsky:2002ni}. The corresponding diagrams lead to terms that scale with powers of \Nc larger than one, in contradiction to the general analysis that the potential is at most of order \Nc.
The authors of Ref.~\cite{Belitsky:2002ni} call this the ``large-\Nc nuclear potential puzzle.''
A possible resolution of this puzzle is proposed in Ref.~\cite{Cohen:2002im} based on the observation that energy-dependent and energy-independent potentials can have different large-\Nc behavior.
The inconsistencies found from multimeson-exchange diagrams all corresponded to energy-dependent potentials. Therefore, Ref.~\cite{Cohen:2002im} suggests that the large-\Nc scalings only apply to energy-independent potentials.
As an example, a class of three-meson exchange contributions to an energy-independent potential are shown to be consistent with the expected large-\Nc scalings, while the analogous contributions to an energy-dependent potential are not.

\subsection{Approximate symmetries of the \NN interactions}
\label{sec:approxsym}

The suppression of the $C_T$ term in Eq.~\eqref{eq:CST-scaling} means that the $S$-wave interactions at very-low energies can be approximated by the single term $C_S (N^\dagger N) (N^\dagger N)$.
Reference \cite{Kaplan:1995yg} notes that this provides a large-\Nc-based explanation for the emergence of Wigner's approximate \Wigner symmetry \cite{Wigner:1936dx,Wigner:1939zz} when $S$ waves dominate the interactions.
Wigner symmetry refers to an invariance under \Wigner transformations among the four spin and isospin states of the nucleons and should not be confused with the underlying contracted SU(4) large-\Nc quark-level symmetry.
The extension of the large-\Nc analysis to the parametrization of the potential in Eq.~\eqref{eq:V_general} shows that, more generally, the \LONc contributions to the central potential are invariant under \Wigner in even partial waves, but that Wigner symmetry is broken by the tensor interaction and in all odd partial waves \cite{Kaplan:1996rk}.
Wigner symmetry in even and odd partial waves was further studied as a so-called long-distance symmetry in Ref.~\cite{CalleCordon:2008cz}, which found agreement with the large-\Nc observations of Ref.~\cite{Kaplan:1996rk}.
However, just as the large-\Nc hierarchy is only manifest in \eftnopi for a range of renormalization point values, Ref.~\cite{Timoteo:2011tt} shows that similarity-renormalization-group (SRG)-evolved \NN interactions exhibit  the long-distance Wigner symmetry only for SRG cutoff values around $\lambda\approx \SI{600}{MeV}$.
The situation is more complex in odd partial waves, for which Ref.~\cite{CalleCordon:2009ps} discusses a long-distance Serber symmetry.\footnote{Serber symmetry refers to the observation that the low-energy $pp$ and $np$ differential cross sections are approximately symmetric about the COM angle $\pi$, see Ref.~\cite{CalleCordon:2009ps} and references therein.}
While the large-\Nc analysis does not necessarily rule out Serber symmetry in the odd partial waves, it does not provide an explanation for it either.

The SU(4)-symmetric Lagrangian of Eq.~\eqref{eq:SU4Lag} motivated the proposal of a so-called ``flip" discrete symmetry by M.~Wise in 2016.\footnote{One of the authors, RPS, gave a talk on the application of this symmetry to few-nucleon systems at the May 9, 2017  Michigan State NSCL seminar on ``Large N constraints on pionless EFT: applications to few bodies;
a window into QCD."}
The term ``flip'' refers to an interchange of quark spin and isospin labels, e.g., a spin-down up quark is transformed into a spin-up down quark. 
Flip symmetry is not a consequence of spin-flavor SU(4), since the flip operation has determinant -1. 
The Lagrangian of Eq.~\eqref{eq:SU4Lag} only describes $S$-wave \NN interactions. 
For these partial waves, the predictions of flip symmetry agree with those of the large-\Nc analysis. 
Flip symmetry does not apply to higher partial waves; e.g., the consequences of imposing flip symmetry to $P$-waves are in disagreement with experiment, and flip symmetry would prohibit \SD mixing.
One interesting aspect of flip symmetry is that it does not mix $N$ and $\Delta$ degrees of freedom in Eq.~\eqref{eq:SU4Lag}, unlike a general SU(4) transformation.
This potentially provides an avenue for understanding why predictions of \NN processes without the $\Delta$ as an intermediate state seem to work well.

Similarly, Ref.~\cite{Lee:2020esp} argues that \NN interactions exhibit a symmetry under the exchange of spin and isospin quantum numbers.
This symmetry is again motivated from the large-\Nc limit.
The \LONc central interactions are indeed invariant under this symmetry but, as found in the flip analysis, it is broken by the \LONc tensor interactions. 
The spin-isospin exchange symmetry is therefore only manifest when tensor interaction effects are removed by considering specific averages over angular momentum channels; a similar approach to that of  Ref.~\cite{CalleCordon:2008cz}.
The authors of Ref.~\cite{Lee:2020esp} further argue that the spin-isospin symmetry is only manifest if the momentum resolution scale is roughly \SI{500}{MeV}, similar to the Wigner symmetry SRG result of Ref.~\cite{Timoteo:2011tt}. 
This is again reminiscent of the need to choose an appropriate value of $\mu$ in \eftnopi for the large-\Nc relationships to hold.

\subsection{Deuteron binding energy}

The previous results show that the expected large-\Nc hierarchies are approximately reflected in available \NN data.
However, one open issue is whether the scaling of the interactions leads to specific large-\Nc scalings of observables. For example, Refs.~\cite{Gross:2011ve,Cohen:2013tya} argue that the deuteron binding energy $B\approx \SI{2.2}{MeV}$ should be of order \Nc.
On the other hand, the $\Delta$-nucleon mass splitting $m_\Delta - m \approx \SI{290}{MeV}$ is of order $1/\Nc$.
This indicates that additional scales can significantly impact the size of physical quantities at $\Nc=3$.
While the binding energy at the physical value of $\Nc=3$ may be explained by certain cancellations, it is not clear whether these cancellations would also occur for other values of \Nc. 
However, other large-\Nc scalings that are consistent with a small value of the binding energy can be found in the literature. 
For example, Ref.~\cite{Beane:2002ab} finds $B\sim \Nc^{-5}$, while Ref.~\cite{Chen:2017tgv} obtains $B\sim \Nc^{-1}$. 
The last result is based on an unconventional scaling of the nucleon axial coupling $g_A\sim \Nc^0$; see Sec.~\ref{sec:nuclear} for further discussion.

\section{SYMMETRY-VIOLATING NUCLEON-NUCLEON INTERACTIONS}

In the symmetry-preserving sector, at least for the lower partial waves, the LECs of the \NN interactions can be determined from the available large and precise dataset on \NN scattering and deuteron properties. 
The large-\Nc analysis provides at least a partial explanation for the observed pattern in the relative magnitudes of the LECs.
With confidence gained from the strong sector, we consider the large-\Nc analysis to be promising in cases for which data is not sufficient to precisely determine all relevant LECs.
A large-\Nc hierarchy can help focus experimental, computational, and theoretical efforts on the most relevant parameters and observables. 
The symmetry-violating \NN interactions provide important examples.
We consider parity-violating and simultaneously time-reversal-conserving (PV); simultaneously time-reversal-violating and parity-violating  (TVPV); and simultaneously time-reversal-violating but parity-conserving   (TVPC) interactions below.

\subsection{Parity-violating and time-reversal-conserving interactions}

Parity violation (PV) in nuclear systems stems from electroweak interactions between quarks in the nucleus, which are dictated by the well-verified electroweak interactions between quarks in the SM.
At nuclear scales, the strong interactions are nonperturbative, which makes it difficult to describe PV in nuclear systems in terms of PV at the quark level.
Conversely, PV in quark interactions can be used to illuminate the impact of the strong interaction in hadronic systems.
For recent reviews of hadronic PV see, e.g., Refs.~\cite{Haxton:2013aca,Schindler:2013yua}.

PV effects in \NN interactions are expected to be suppressed by a factor of $\approx$ $10^{-7}$ compared to parity-conserving (PC) terms.
This suppression can make experimental determinations of PV effects in nuclear systems challenging. 
Exceptions exist for some heavier nuclei, in which enhancements of several orders of magnitude have been observed; see, e.g., Ref.~\cite{Bowman:1989ci}. 
However, the theoretical description of these heavier systems in terms of the underlying \NN interactions is currently not well understood.
In light nuclei, for which a  theoretical description in terms of two- and few-nucleon interactions is feasible, PV effects are not enhanced.
The number of experiments that have observed PV in light nuclei is therefore  limited; an accurate and  reliable extraction of the PV parameters in the \NN interactions from such experiments is still lacking.
In principle, the PV parameters can be determined from the underlying quark interactions using lattice QCD, but efforts in this direction are  still in the  nascent stage \cite{Wasem:2011tp,Kurth:2015cvl}.
In the absence of data and precise lattice QCD determinations, the large-\Nc expansion is useful for obtaining theoretical constraints on the relative sizes of the PV parameters.

\subsubsection{General parametrization of the PV interactions}
The general form of the PV potential up to NNLO in the 1/\Nc expansion was presented in Ref.~\cite{Phillips:2014kna}.
The \LONc interactions are $O(\Nc)$ and contain isoscalar and isotensor terms,
\begin{align}\label{eq:PV-PSS-LO}
    & \pminus \cdot \left( \vsig_1 \times \vsig_2 \right) \vtau_1 \cdot \vtau_2 \, ,
    &&
     \pminus \cdot \left( \vsig_1 \times \vsig_2 \right) \calI_{ab} \tau_1^a \tau_2^b \, ,
\end{align}
where $\calI_{ab} = \text{diag}(1,1,-2)$.
Unlike in the symmetry-preserving case, the \NLONc PV \NN contributions are $O(\Nc^0)$ and are suppressed only by a single factor of 1/\Nc.
The \NLONc potential contains only isovector terms, 
\begin{equation}\label{eq:PV-PSS-NLO}
\begin{aligned}[c]
&\pplus \cdot \left( \vsig_1 \tau_1^z - \vsig_2 \tau_2^z \right) \, ,\\
     &\pminus \cdot \left( \vsig_1 + \vsig_2 \right) [\vtau_1 \times \vtau_2 ]^z\, ,
\end{aligned}
\hspace{8em}
\begin{aligned}[c]
& \pminus \cdot \left( \vsig_1 \times \vsig_2 \right) \left[ \vtau_1 + \vtau_2 \right]^z \, , \\
    & \! \![ \left( \pplus \times \pminus \right)^i p_-^j ]_2  \cdot [ \sigma^i_1 \sigma^j_2 ]_2 [\vtau_1 \times \vtau_2 ]^z\, ,
\end{aligned}
\end{equation}
where the tensor structure  $[uv]_2^{ij} \equiv u^i v^j + u^j v^i - \frac{2}{3} \vec{u} \cdot \vec{v}$.
The \NNLONc terms are suppressed by one additional power of 1/\Nc and contain isoscalar and isotensor contributions,
\begin{equation}\label{eq:PV-PSS-NNLO}
\begin{split}
\begin{aligned}[c]
    & \pminus \cdot\,(\vsig_1\times \vsig_2) \, , \\
    & \pplus^{\,2}\,\pminus\cdot\,(\vsig_1\times \vsig_2)\,\vtau_1\cdot\,\vtau_2\, , \\
    & \pplus\cdot\,(\vsig_1 - \vsig_2)\, , 
    \end{aligned}
\hspace{8em}
\begin{aligned}[c]
    & \pplus\cdot\,(\vsig_1 - \vsig_2) \, \vtau_1\cdot\,\vtau_2\, , \\
    & \pplus\cdot\, (\vsig_1 -
        \vsig_2) \,\calI_{ab} \tau_1^a \tau_2^b\, , \\
    & \pplus^{\,2}\,\pminus\cdot\,(\vsig_1\times
        \vsig_2) \,\calI_{ab} \tau_1^a \tau_2^b\, .
        \end{aligned}
\end{split}
\end{equation}
The PV interactions highlight the impact of the momentum scaling rules of Eq.~\eqref{eq:MomScale}.
The spin-isospin structures of the first terms in Eqs.~\eqref{eq:PV-PSS-LO} and \eqref{eq:PV-PSS-NLO} are proportional to combinations of $G^{ia} G^{jb}$ and $G^{ia} \1$, respectively, thus leading to the same large-\Nc scaling.
The suppression of the term in Eq.~\eqref{eq:PV-PSS-NLO} comes from the factor of $1/\Nc$ accompanying the momentum $\pplus$ relative to $\pminus$ in the term of Eq.~\eqref{eq:PV-PSS-LO}.

In Ref.~\cite{Phillips:2014kna}, these operator forms are multiplied by arbitrary functions of $\pminus^{\,2}$, which do not impact the large-\Nc scaling.
In the general PV potential the suppression from one order to the next is 1/\Nc; however, in a given isospin sector the suppression is in powers of 1/$\Nc^2$, as observed in the symmetry-preserving case.
At the weak matching scale, the isovector and isotensor terms include an additional factor of $\sin^2\theta_W \approx 0.23$ \cite{ParticleDataGroup:2022pth}, comparable to an additional suppression of 1/\Nc for the physical value of $\Nc=3$ and Ref.~\cite{Phillips:2014kna} advocates for including it in the scaling of the operators at $\Nc=3$.
However, even though the running of the quark-level operators is understood in perturbative QCD \cite{Dai:1991bx,Tiburzi:2012hx,Tiburzi:2012xx,Gardner:2022mxf,Gardner:2022dwi} (that is, down to $\approx$ 1-2 GeV), it is currently not known how this factor is modified by the nonperturbative evolution of the LECs to the hadronic scale.
    
\subsubsection{Impact on phenomenological models}
Reference \cite{Phillips:2014kna} mapped the results of their analysis onto the so-called DDH potential \cite{Desplanques:1979hn}.
In this phenomenological approach, the PV \NN interactions are parametrized in terms of single-meson exchanges. 
The long-range PV interaction is given by one-pion exchange, while single $\rho$ and $\omega$ meson exchanges serve as parametrizations of short-range \NN interactions.
However, the single-meson exchanges of the DDH model do not generate the spin tensor structure in Eq.~\eqref{eq:PV-PSS-NLO}.
Reference \cite{Desplanques:1979hn} estimated ``best values'' and ``reasonable ranges'' for their PV meson-nucleon couplings based on a quark model, symmetry considerations, and the factorization approximation.
The large-\Nc hierarchy obtained by Ref.~\cite{Phillips:2014kna} approximately matches the hierarchy of the estimated ``best values'' of the PV meson-nucleon couplings in the DDH potential, with the important exception of the PV pion-nucleon coupling $h_\pi^1$, for which the ``best value'' is much larger than the large-\Nc expectation.
However, the ``reasonable ranges'' of the PV meson-nucleon couplings given in \cite{Desplanques:1979hn} are too large to draw any firm conclusions.
Reference \cite{Phillips:2014kna} finds the PV pion-nucleon coupling to scale as $h_\pi^1 \lesssim 1/\Nc^{-1/2}$.
This is in agreement with the result of Ref.~\cite{Zhu:2009nj}, in which PV meson-nucleon scattering was analyzed and consistency conditions between pion-nucleon and pion-$\Delta$ coefficients were derived in analogy to the PC case \cite{Gervais:1983wq, Gervais:1984rc,Dashen:1993ac, Dashen:1993as}.

\subsubsection{Combined large-\Nc and \eftnopi expansion}
The terms in Eqs.~\eqref{eq:PV-PSS-LO}-\eqref{eq:PV-PSS-NNLO} give the most general form of the PV potential through \NNLONc, without any additional hierarchies. 
Such additional hierarchies exist in EFT approaches, and the large-\Nc expansion is particularly useful when combined with an EFT expansion.
This approach was taken in Ref.~\cite{Schindler:2015nga}, analyzing the large-\Nc dependence of the LO terms in PV \eftnopi.
In \eftnopi, the LO (in the EFT power counting) PV Lagrangian consists of five terms describing \SP transitions \cite{Zhu:2004vw,Girlanda:2008ts,Phillips:2008hn,Schindler:2009wd}.\footnote{These represent the field theoretic generalization of the Danilov amplitudes \cite{Danilov:1965,Danilov:1971fh}.}
The corresponding LECs in an operator basis that makes the partial-wave structure explicit are denoted by $\threeSoneP$, $\oneSthreePscalar$, $\oneSthreePvector$, $\oneSthreePtensor$, and $\threeSthreeP$.
(As discussed in detail in Ref.~\cite{Schindler:2015nga}, the application of a na\"ive large-\Nc analysis to the minimal form of the Lagrangian of Ref.~\cite{Girlanda:2008ts} results in incorrect scaling assignments.)
In a combined \eftnopi and large-\Nc expansion, there are two independent terms at \LONc: the isotensor term $\oneSthreePtensor$ and one isoscalar term. 
Both the isoscalar LECs $\threeSoneP$ and $\oneSthreePscalar$ are \LONc, but they are related to each other by
\begin{align}
\label{eq:PVisoscalarId}
    \threeSoneP = 3 \, \oneSthreePscalar \,
\end{align}
up to corrections of order $1/\Nc^2$.
Isovector PV \NN interactions only appear at \NLONc.

At three derivatives in PV \eftnopi, an additional 11 operators contribute to PV elastic \NN scattering, consisting of corrections to the \SP coefficients and the first contributions to the six allowed \PD transitions.
The large-\Nc analysis of these terms was performed in Ref.~\cite{Nguyen:2020quk}.
The results for these LECs follow the pattern found at LO in the \eftnopi power counting; only two independent three-derivative LECs are \LONc, one isotensor and one isoscalar, and two isoscalar terms are related to each other up to corrections of order $1/\Nc^2$.

\subsubsection{Comparison with experiment}
These large-\Nc results conflict with the meson-exchange picture used to analyze existing PV data  \cite{Desplanques:1979hn}.
The latter approach assumed that PV pion exchange, proportional to the isovector coupling $h_\pi^1$, plays a major role.
Results are often displayed in a two-dimensional projection \cite{Adelberger:1985ik} onto specific linear combinations of isoscalar and isovector PV meson-nucleon couplings by letting the isotensor contributions vary over a specific range that was based on the estimated ``best values'' from Ref.~\cite{Desplanques:1979hn}.
However, this analysis leads to some tension when combined with the results from different experiments, see the discussion in Ref.~\cite{Gardner:2017xyl}.
One explanation is that the isoscalar and isovector axes contribute at different orders in the large-\Nc expansion; an appropriate theoretical error analysis needs to reflect that.  
Instead, the dominant interactions are the isoscalar and isotensor interactions.

Several experiments have been performed in the two-nucleon sector that can be used to test the large-\Nc expectations.
It is important to keep in mind that the expected suppression of the \NLONc contribution is only by a single factor of $1/\Nc$, unlike in the symmetry-preserving interactions.
This means that ``natural'' factors of order 1 can obscure the relative large-\Nc scaling of two LECs at $\Nc=3$.

The longitudinal asymmetry $A_L^{pp}$ in proton-proton scattering at \SI{13.6}{MeV} \cite{Eversheim:1991tg,Haeberli:1995uz} constrains a linear combination of isoscalar, isovector, and isotensor LECs \cite{Phillips:2008hn}. 
The upper bound on the induced circular photon polarization $P_\gamma$ in unpolarized neutron capture on protons \cite{Knyazkov:1984ke,Knyazkov:1984zz} provides a constraint on the other isoscalar LEC and the isotensor LEC \cite{Schindler:2009wd}.
If the large-\Nc analysis is valid, the isovector contribution to $A_L^{pp}$ can be neglected at \LONc. 
By assuming the validity of the relationship in Eq.~\eqref{eq:PVisoscalarId}, the two observables depend on two unknown LECs.
The observables are  proportional to ratios of PV to PC LECs. 
As discussed in Ref.~\cite{Phillips:2008hn,Schindler:2009wd}, the renormalization point dependence of the 
PV LECs is given by that of the corresponding $S$-wave PC LECs;
therefore, the PV to PC ratios of LECs are renormalization point independent.
As discussed in Sec.~\ref{sec:sym_preserve}, the two $S$-wave LECs are expected to be identical in the large-\Nc limit.
In the following we denote the corresponding strong LEC simply by $C$ without reference to a particular partial wave.
Extracting the ratios in the isoscalar and isotensor sectors from the experimental results for $A_L^{pp}$ and $P_\gamma$ yields
\begin{align}
\label{eq:PVisoscalartensornumbers}
    \threeSoneP/C = 
    \SI{-1.1 \pm 1.0e-10}{MeV}^{-1}
    \, , \quad 
    \oneSthreePtensor/C = \SI{7.4 \pm 2.3e-11}{MeV}^{-1}\, .
\end{align}
The large uncertainty on the isoscalar LEC reflects the large uncertainty on the upper bound on $P_\gamma$.

A test of the large-\Nc hierarchy also requires knowledge of the sizes of the \NLONc LECs.
The $\mu$-independent ratio of the \NLONc isovector LEC \threeSthreeP with the strong interaction LEC can be determined from the PV asymmetry $A_\gamma$ in polarized neutron capture, $\vec{n}p\to d\gamma$ \cite{NPDGamma:2018vhh}. 
At LO and NLO in the \eftnopi power counting, there is no \LONc contribution to $A_\gamma$ \cite{Savage:2000iv,Schindler:2009wd}. 
The extracted ratio is\footnote{This result differs from the extraction that appears in Ref.~\cite{NPDGamma:2018vhh}.}
\begin{align}
\label{eq:PVisovectornumber}
    \threeSthreeP/C = \SI{-4.6\pm 2.1e-11}{MeV}^{-1}\, .
\end{align}
Comparing the sizes of the \LONc ratios in Eq.~\eqref{eq:PVisoscalartensornumbers} with the \NLONc ratio in Eq.~\eqref{eq:PVisovectornumber} shows that, while the uncertainties are large, the central values are not inconsistent with the predicted large-\Nc hierarchy. 

Some authors have argued that a combined analysis of existing experiments indicates that the isovector PV interaction in the $\threeS$-$\Pone$ channel might not be suppressed  \cite{Vanasse:2018buq,Haxton:cipanp18}.  
For example, Ref.~\cite{Vanasse:2018buq} determines a linear combination of \LONc isoscalar LECs from an \eftnopi calculation of the PV longitudinal asymmetry $A_L^{pd}$ in proton-deuteron scattering, and uses that result in the \eftnopi expression for $A_L^{pp}$ to determine the isotensor LEC.
The obtained isoscalar and isotensor LECs are shown to give a value for $P_\gamma$ consistent with the experimental bound  and they are also of the same size as the isovector LEC obtained from $A_\gamma$, casting doubt on the utility of the large-\Nc analysis. 
However, the experimental result for the proton-deuteron asymmetry, $A_L^{pd}(E_\text{lab}=\SI{15}{MeV}) = (-3.5 \pm 8.5) \times 10^{-8}$ has large uncertainties.
Using the values given in Eq.~\eqref{eq:PVisoscalartensornumbers} yields $A_L^{pd}(E_\text{lab}=\SI{15}{MeV}) = 1.5\times 10^{-7} $,  not inconsistent with the experimental bound.
Further, using the identity in Eq.~\eqref{eq:PVisoscalarId}, the isoscalar linear combination determined in Ref.~\cite{Vanasse:2018buq} is $\frac{1}{2} \threeSoneP/C$. 
Given that the suppression of the isovector LEC is only by a single factor of $1/\Nc = 1/3$, 
the dual expansion expectation is vulnerable to a variety of ${\cal O}(1)$ effects.  This can include accidental numerical factors, a change in basis as described in Sec 3, etc. What can be said at the moment is that, with an appropriate treatment of experimental and theoretical errors, the dual expansion yields results that are not inconsistent with available data.

Comparisons with two- and three-body experiments can be performed consistently using \eftnopi.
Since data from these systems is limited and at least in some cases comes with large uncertainties, Ref.~\cite{Gardner:2017xyl} extends the analysis of the large-\Nc results to include data beyond three-nucleon systems, such as the recently measured asymmetry in neutron capture on ${}^3\text{He}$ \cite{n3He:2020zwd} and the circular photon polarization from ${}^{18}\text{F}$ decay \cite{Barnes:1978sq,Ahrens:1982vfn,Bini:1985zz,Page:1987ak}.
Overall, Ref.~\cite{Gardner:2017xyl} finds good agreement with the large-\Nc expectations.
However, the analysis of Ref.~\cite{Gardner:2017xyl} was performed before the result of Ref.~\cite{NPDGamma:2018vhh} for $A_\gamma$ was published.
An updated analysis \cite{Haxton:cipanp18} that includes the $A_\gamma$ result casts doubt on this agreement, since the resulting isovector parameter  appears to not be suppressed, similar to the conclusion of Ref.~\cite{Vanasse:2018buq} discussed above.
Both the seeming agreement of Ref.~\cite{Gardner:2017xyl} and the seeming disagreement of Ref.~\cite{Haxton:cipanp18} should be interpreted with the caveats that come from attempting to apply an EFT outside its realm of applicability, as well as using LECs from one theory in another. References~\cite{Gardner:2017xyl,Haxton:cipanp18} are based on calculations of PV observables in a variety of formalisms.
The authors relate the PV parameters in each formalism to one common set of parameters by matching at very low energies.
One problem with this approach is that the matching between different formalisms introduces implicit (and unknown) scale dependence \cite{Schindler:2013yua}, but the parameters used in Refs.~\cite{Gardner:2017xyl,Haxton:cipanp18} are taken to be scale-independent.
As demonstrated in Sec.~\ref{sec:sym_preserve}, the large-\Nc analysis constrains LECs, which are not observables, and the agreement of the large-\Nc hierarchies with data sensitively depends on the choice of renormalization scale.
It is therefore important to perform consistent calculations in which the renormalization dependence can be eliminated or at least minimized.

\subsection{Time-reversal-violating and parity-violating interactions}
    
The search for TVPV effects is an important component of the search for BSM physics.
Although the SM contains sources of TVPV, such as the phases of the CKM matrix and the $\theta$ term of QCD, their effects are predicted to be much smaller than any current experimental limits.
An observation of TVPV effects larger than those allowed in the SM would be a signal of BSM physics.
As in the case of PV, the symmetry-violating interactions between two and possibly more nucleons becomes important for processes involving nuclei.
Several approaches, including general parametrizations \cite{Herczeg:1966}, meson-exchange models (see, e.g., Refs.~\cite{Simonius:1975ve,Haxton:1983dq,Gudkov:1992yc,Towner:1994qe,Liu:2004tq}), and EFTs \cite{Mereghetti:2010tp,Maekawa:2011vs,deVries:2012ab,Maekawa:2011vs,Bsaisou:2012rg,Bsaisou:2014zwa,Bsaisou:2014oka} (see Ref.~\cite{deVries:2020iea} for a review) have been considered. 
Because of the lack of data, the corresponding TVPV parameters, such as TVPV meson-nucleon couplings or TVPV \NN LECs, are currently not well constrained. 
The large-\Nc analysis of TVPV interactions can provide theoretical guidance on the relative sizes of the multitude of TVPV parameters.

\subsubsection{General parametrization}
In the TVPV case, there is a single operator structure at \LONc \cite{Samart:2016ufg}, 
\begin{equation}
    \pminus \cdot(\vec\sigma_1 \tau_1^z - \vec\sigma_2 \tau_2^z) \ ,
\end{equation}
which is $\calO(\Nc)$. This operator is an isovector term. At \NLONc, $\calO(\Nc^0)$, Ref.~\cite{Samart:2016ufg} lists five additional terms,
\begin{align}
    & \pminus \cdot (\vec\sigma_1 - \vec\sigma_2) && {}
     \nonumber \\
    & \pminus \cdot (\vec\sigma_1 - \vec\sigma_2) \vec\tau_1\cdot \vec\tau_2
    && \pminus \cdot (\vec\sigma_1 - \vec\sigma_2) \calI_{ab}\tau_1^a\tau_2^b
     \\
    & \pplus \cdot (\vec\sigma_1 \times \vec\sigma_2)\vec\tau_1\cdot \vec\tau_2 
    && \pplus \cdot (\vec\sigma_1 \times \vec\sigma_2) \calI_{ab}\tau_1^a\tau_2^b \nonumber
\end{align}
These are either isoscalar (left-hand column) or isotensor (right-hand column) terms. 
As discussed below, not all of these terms are independent when considered as originating from a Lagrangian with a single derivative.
The complete list of terms at \NNLONc can be found in Ref.~\cite{Samart:2016ufg}. 
These are again isovector operators, so that while the expansion of the complete interaction is in 1/\Nc, for a given isospin sector the expansion is in $1/\Nc^2$, analogous to the symmetry-conserving and PV interactions. 

\subsubsection{Impact on phenomenological models}
These general operator structures can be matched onto particular TVPV \NN interactions, including meson-exchange models and EFTs. 
In the single-meson-exchange picture, the \LONc contribution is parametrized by isovector pion and $\omega$ exchanges \cite{Liu:2004tq}.
Of particular interest in the meson-exchange picture are the TVPV pion-nucleon couplings, which have analogues in ChEFT. 
The analysis of Ref.~\cite{Samart:2016ufg} shows that among these terms, the isovector $\pi N$ coupling is dominant in the large-\Nc expansion, $\bar{g}_\pi^{(1)} \sim O(\Nc^{1/2})$, while the isoscalar $\bar{g}_\pi^{(0)}$ and isotensor $\bar{g}_\pi^{(2)}$ parameters are both $O(\Nc^{-1/2})$; suppressed by a factor of 1/\Nc.

\subsubsection{Impact on EFTs}
Reference~\cite{Samart:2016ufg} considered the general operator structures at a given order in the large-\Nc expansion. Each term listed in that work can be multiplied by functions of $\pminus^{\; 2}$ because these do not change the large-\Nc scaling. 
One can also interpret the operators as originating from \eftnopi by replacing the arbitrary $\pminus^{\; 2}$-dependent functions by LECs. 
However, as in the T-even PV case, the number of terms at a given order in the large-\Nc expansion can be reduced by the use of Fierz identities.
For the TVPV case, this analysis is performed in Ref.~\cite{Vanasse:2019fzl}. The authors show that there are only three independent operators at \NLONc and one operator at \NNLONc when considering terms with a single factor of momentum.

In ChEFT, the LO (in the chiral power counting) potential consists of two \NN contact terms as well as pion exchange contributions with one isoscalar, one isovector, and one isotensor $\pi N$ LEC \cite{Mereghetti:2010tp,Maekawa:2011vs,deVries:2012ab,Maekawa:2011vs,Bsaisou:2012rg,Bsaisou:2014zwa,Bsaisou:2014oka}.
Only the pion-exchange contribution proportional to the isovector LEC $\bar{g}_\pi^{(1)}$ is \LONc, with the other terms being suppressed by a factor of 1/\Nc. 
This suggests that the isovector LEC $\bar{g}_\pi^{(1)}$ should be of particular interest to future lattice QCD studies (see Ref.~\cite{deVries:2016jox} for a proposed strategy of determining $\bar{g}_\pi^{(1)}$ from spectroscopy).
However, as discussed in Ref.~\cite{Samart:2016ufg}, given the physical value $1/\Nc = 1/3$ and that LECs that are considered natural can vary by factors of order unity, it seems reasonable to retain all terms that are LO in the chiral power counting but emphasize those that are also leading in large-\Nc when calculating observables. 

\subsection{Time-reversal-violating and parity-conserving  interactions}

While the SM contains both PV and TVPV interactions, TVPC effects can only arise in the SM through interference of these two highly suppressed interactions. On the other hand, BSM physics may include direct TVPC sources.
The TVPC \NN interactions are analyzed in Ref.~\cite{Samart:2016ufg}. There are two terms at \LONc, 
\begin{align}
\label{eq:TVPC}
    &p_-^{i} p_+^{\,j} [\sigma_1 \sigma_2]_2^{ij} \vec\tau_1 \cdot \vec\tau_2\ , &
    &p_-^{i} p_+^{\,j} [\sigma_1 \sigma_2]_2^{ij} \calI_{ab}\tau_1^a\tau_2^b\ ,
\end{align}
with one isoscalar and one isotensor term. 
(The spin tensor structure is defined below Eq.~\eqref{eq:PV-PSS-NLO}.)
The \NLONc TVPC terms are suppressed by a single factor of 1/\Nc relative to the \LONc TVPC terms and are all isovector, while the \NNLONc terms are suppressed by another factor of 1/\Nc and are isoscalar and isotensor. 
There is thus again an expansion in $1/\Nc^2$ when considering a specific isospin sector. 

TVPC interactions have received less phenomenological attention than TVPC physics. 
A general parametrization of the TVPC potential was derived in Ref.~\cite{Herczeg:1966}, but most phenomenogical applications have considered meson-exchange models.
Single-pion-exchange cannot contribute to TVPC interactions \cite{Simonius:1975ve}. 
The lightest mesons that can contribute to the TVPC single-meson-exchange potential are the $\rho(770)$ and the $h_1(1170)$ (see, e.g., Ref.~\cite{Song:2011jh}). 
However, $\rho$ exchange corresponds to a \NLONc operator, while $h_1(1170)$ matches onto a \NNLONc term.
In other words, \LONc terms are not generated by only including these two mesons.
The isoscalar \LONc operator structure \emph {can} be generated by the exchange of an $a_1(1260)$ meson \cite{Song:2011jh}.
This suggests that $a_1(1260)$ be included in meson-exchange treatments \cite{Samart:2016ufg}. 

Because of the absence of one-pion-exchange \cite{Simonius:1975ve}, contact interactions with two derivatives are expected to provide the leading contribution to TVPC \NN interactions in the EFT framework. 
Both terms in Eq.~\eqref{eq:TVPC} correspond to two-derivative contact interactions, na\"ively  contributing at the same order in the EFT power counting. 
However, Ref.~\cite{Simonius:1975ve} shows that the lowest partial-wave transition that can receive isotensor contributions is $^3\!P_2$-$^3\!F_2$.
An example is the isotensor structure in Eq.~\eqref{eq:TVPC} multiplied by a function of $p^2$, corresponding to an operator with four derivatives, which is suppressed in the EFT power counting.
Alternatively, at the Lagrangian level in the center-of-mass frame, the isotensor operator in Eq.~\eqref{eq:TVPC} vanishes from Fierz transformations \cite{Inoue:comm}.
Therefore, in a combined EFT and large-\Nc expansion, only the isoscalar operator in Eq.~\eqref{eq:TVPC} contributes.

\section{EXTERNAL CURRENTS}\label{sec:currents}

Including external currents in the large-\Nc analysis will allow us to order two-and three-nucleon interactions that involve electroweak processes as well as possible BSM physics.
Electroweak probes provide a sensitive test of SM physics occurring within nuclei, and many BSM searches involve nuclei, so we wish to generalize the large-\Nc counting rules developed for matrix elements of the type in Eq.~\eqref{matching} to include a non-baryonic external field $A$ in the incoming or outgoing state,
\begin{align}
    \langle N N A | H | N N \rangle \, .
\end{align}
The external field may also carry spin and isospin indices. 
After contraction of the field in the Hamiltonian and the external state, the remaining baryon matrix elements are equivalent to those in the purely hadronic case and the same methods can be employed to determine the large-\Nc scaling.

In order to study electroweak processes, a choice must be made regarding the scaling of the nucleon charge with \Nc. 
One option is to keep the quark charges fixed at their physical values, in which the case the nucleon charge grows with \Nc and the neutron is no longer electrically neutral.
The other option is to allow the quark charges to be \Nc-dependent such that the nucleon charge is fixed at the physical value and independent of \Nc.
The latter is consistent with anomaly cancellations in the Standard Model \cite{Chow:1995by,Shrock:2002kp}, and this is the choice we make in the following (for more details on the charge scaling see App.~A in Ref.~\cite{Richardson:2021xiu}).

For current operators that arise from gauging derivatives, the large-\Nc scaling of the LECs can be determined from the corresponding \NN interaction terms without external fields.
The electromagnetic minimally-coupled case is discussed in Refs.~\cite{Riska:2002vn,Richardson:2020iqi}, for which there are two independent operator structures at both \LONc and \NNLONc. 
In Ref.~\cite{Richardson:2020iqi} the large-\Nc scalings of the LECs are determined from the two-derivative \NN interactions of Ref.~\cite{Schindler:2018irz} (see Sec.~\ref{sec:twoderivative}).
These results agree with the scalings found in Ref.~\cite{Riska:2002vn}.
In the following, we focus on current operators that do not arise from gauging derivatives.

\subsection{Magnetic and axial two-body currents}\label{sec:magnetic}

At LO in \eftnopi, there are two independent operators coupling a two-nucleon system to an external magnetic field. In the partial wave basis the corresponding LECs are $\Lone$ and $\Ltwo$ \cite{Chen:1999tn},
\begin{align}
\label{external:EM:lagr_pw}
    \calL = e B_i \left[ \Lone \left( N^T P_i N \right)^\dagger \left( N^T \bar{P}_3 N \right) - i \epsilon^{ijk} \, \Ltwo \left( N^T P_j N \right)^\dagger \left( N^T P_k N \right)\right] +\text{h.c.} \, ,
\end{align}
where $P_i = \frac{1}{\sqrt{8}} \sigma_2\sigma_i\tau_2$ and $\bar{P}_3 = \frac{1}{\sqrt{8}} \sigma_2\tau_2\tau_3$ project onto $\threeS$ and $\oneS$ partial waves, respectively.
Fits to radiative neutron capture and the deuteron magnetic moment determine the LECs \Lone and \Ltwo, respectively \cite{Chen:1999tn}:
\begin{equation}
    \label{eq:mag_exp_values}
    \Lone (\mu = m_\pi) = 7.24 \, \text{fm}^4 \, , \quad \Ltwo (\mu = m_\pi) = -0.149 \, \text{fm}^4 \, .
\end{equation}
Despite contributing at the same order in the \eftnopi power counting, the values of the two LECs are very different.
The large-\Nc analysis of Ref.~\cite{Richardson:2020iqi} provides a possible explanation for this difference.
Applying the rules of Sec.~\ref{sec:aspects}, we start with the most general form of the Lagrangian in the large-\Nc basis and perform Fierz transformations while keeping the leading large-\Nc scaling for each term explicit.
This leads to 
\begin{align}
\label{external:EM:min_lagr}
    \calL = e B^i \left[ \CMs \left( N^\dagger \sigma^i N \right) \left( N^\dagger N \right)
    +\CMv \epsilon^{ijk} \epsilon^{3ab} \left( N^\dagger \sigma^j \tau^a N \right) \left( N^\dagger \sigma^k \tau^b N \right) \right] \, ,
\end{align}
with the large-\Nc scalings 
\begin{align}
        \CMs  \sim O(1) \, , \quad \CMv  \sim O(\Nc) \,.
    \end{align}
The large-\Nc basis LECs are related to the partial-wave basis LECs by
\begin{align}
\label{external:EM:Chen_relation}
    \Lone = 8 \CMv , \quad \Ltwo = - \CMs \, .
\end{align}
Using the values of Eq.~\eqref{eq:mag_exp_values} gives
\begin{align}
\CMs(\mu = m_\pi) = \SI{0.149}{fm}^4\, , \quad \CMv(\mu = m_\pi) = \SI{0.905}{fm}^4 \, .
\end{align}
Experimentally, \CMs is suppressed by a factor of 6 compared to \CMv at $\mu=m_\pi$.  This additional factor of 2 compared to $\Nc=3$ is what one might expect given the existence of factors not due to large-\Nc symmetries, and the fact that the LECs are $\mu$-dependent (although the $\mu$-dependence in the ratio is mild). The apparently larger deviation from large-\Nc expectations that occurs when considering the problem in the partial-wave basis is reminiscent of what happens in the \NN $P$-waves (see Sec.~\ref{sec:twoderivative}) where agreement with large-\Nc scaling is better in the large-\Nc basis than in the partial-wave basis. 

This analysis is generalized to arbitrary axial fields in Ref.~\cite{Richardson:2020iqi} by replacing the magnetic field with an axial field in Eq.~\eqref{external:EM:lagr_pw}, with the corresponding axial LECs denoted by $L_{1,A}$ and $L_{2,A}$. 
These operators are relevant for, e.g., neutrino-deuteron scattering.
The same large-\Nc ratio as in the magnetic case holds for the isoscalar and isovector LECs,
    \begin{equation}
    \label{external:axial:ratio}
        \left \vert \frac{L_{2,A}}{L_{1,A}} \right \vert_{\Nc}   \approx \frac{1}{8 \Nc} \, .
    \end{equation}
However, unlike in the magnetic case, only one of the LECs ($L_{1,A}$) has been reliably determined \cite{Butler:1999sv,Butler:2002cw,Chen:2002pv,Ando:2008va,De-Leon:2016wyu,Acharya:2019fij}.
Combining the large-\Nc ratio of Eq.~\eqref{external:axial:ratio} with the result of a lattice calculation by the NPLQCD collaboration \cite{Savage:2016kon} yields
    \begin{equation}
        L_{2,A} \approx 0.1625 \, \text{fm}^3\, ,
    \end{equation}
for the previously undetermined axial LEC at $\Nc=3$.

These results are consistent with the lattice QCD results of Ref.~\cite{Detmold:2021oro}. 
Those authors find allowed regions for each LEC by matching lattice QCD results of multiple two- and three-body observables to \eftnopi calculations performed in the same finite volumes. 
While the formulation of the finite-volume \eftnopi uses a Gaussian regulator function, which prohibits a direct comparison with formulations based on dimensional regularization and PDS, the observed trends are consistent with the expected suppression of \Ltwo relative to \Lone. 
For axial LECs, Ref.~\cite{Detmold:2021oro} determines the ratios $\tilde{L}_{1,A} = \LoneA/g_A$ and $\tilde{L}_{2,A} = \LtwoA/g_{A,0}$, where $g_A$ and $g_{A,0}$ are the isovector and isoscalar nucleon axial charge, respectively.
Since the isoscalar currents in the one- and two-nucleon sector are both suppressed by a factor of 1/\Nc relative to the corresponding isovector currents, the ratios $\tilde{L}_{1,A}$ and $\tilde{L}_{2,A}$ are of the same order in the 1/\Nc expansion. 
While the values extracted in Ref.~\cite{Detmold:2021oro} have relatively large uncertainties and some tension exists for different extractions of $\tilde{L}_{2,A}$, these results do not contradict large-\Nc scaling.

\subsection{Neutrinoless double beta decay and charge independence breaking}
The light Majorana exchange mechanism for neutrinoless double beta decay requires a short-range two-nucleon current (with LEC $\gnu$) at leading order in both \eftnopi and ChEFT  \cite{Cirigliano:2018hja,Cirigliano:2019vdj}.
Chiral symmetry relates this term to charge-independence-breaking (CIB) \NN interactions \cite{Cirigliano:2017tvr},   
\begin{align}
        \calL_{C\!I\!B}^{\NN} & =  \frac{e^2}{4} \left\{ \calC_1 \left[ \left( \bar N u^\dagger \tilde Q_R u N \right)^2 + \left( \bar N u \tilde Q_L u^\dagger N \right)^2 - \frac{1}{6} \Tr( \tilde Q_L^2 + \tilde Q_R^2 ) \left(\bar N \tau^a N \right)^2 \right] \right. \nonumber \\
        &   + \left. \calC_2 \left[  2 \left( \bar N u^\dagger \tilde Q_R u N \right) \left( \bar N u \tilde Q_L u^\dagger N \right) - \frac{1}{3} \Tr( U \tilde Q_L U^\dagger \tilde Q_R ) \left( \bar N \tau^a N \right)^2  \right] \right\} + \cdots, \label{eq:CIB:1}  
\end{align}
where in this case $\tilde Q_L = \tilde Q_R = \frac{1}{2} \tau^3$ and $U = u^2 = \exp[\frac{i}{F}\phi_a \tau^a]$, with $\phi_a$ $(a=1,2,3)$  the pion fields in Cartesian coordinates and $F$ the pion decay constant in the chiral limit.
The ellipse denotes terms that are not relevant for the relationship to neutrinoless double beta decay.
In particular, Ref.~\cite{Cirigliano:2017tvr} showed that $\gnu = \calC_1$.
Currently, only the combination $\calC_1+\calC_2$ can be determined experimentally. 
To estimate the impact of the undetermined LEC $\gnu$, Refs.~\cite{Cirigliano:2018hja,Cirigliano:2019vdj} approximate $\gnu = \frac{1}{2} \left( \calC_1 + \calC_2 \right)$.
This assumption is justified from the large-\Nc perspective in Ref.~\cite{Richardson:2021xiu} by showing that both $\calC_1$ and $\calC_2$ contribute at \LONc and that $\calC_1 = \calC_2 [1+O(1/\Nc)]$; i.e., $\gnu = \frac{1}{2} \left( \calC_1 + \calC_2 \right)$ at LO in the large-\Nc expansion.
These constraints are also in agreement with a recent estimate of $\calC_1$ and $\calC_2$ from dispersion theory \cite{Cirigliano:2020dmx, Cirigliano:2021qko}.
    
\subsection{Interactions with dark matter}

\begin{figure}[t!]
    \centering
    \begin{subfigure}[t]{0.5\textwidth}
        \centering
        \includegraphics[width=\textwidth]{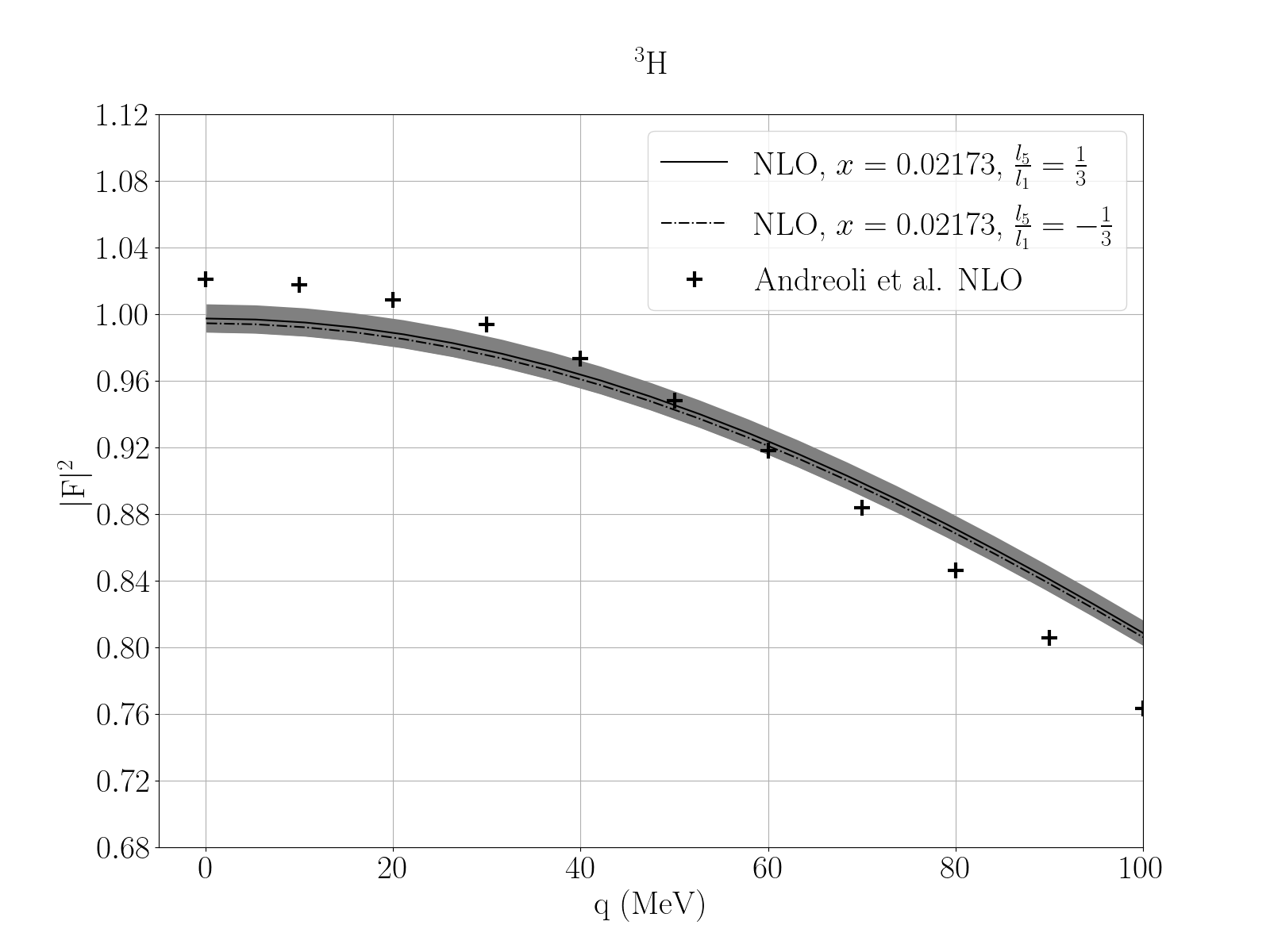}
    \end{subfigure}%
    ~
    \begin{subfigure}[t]{0.5\textwidth}
        \centering
        \includegraphics[width=\textwidth]{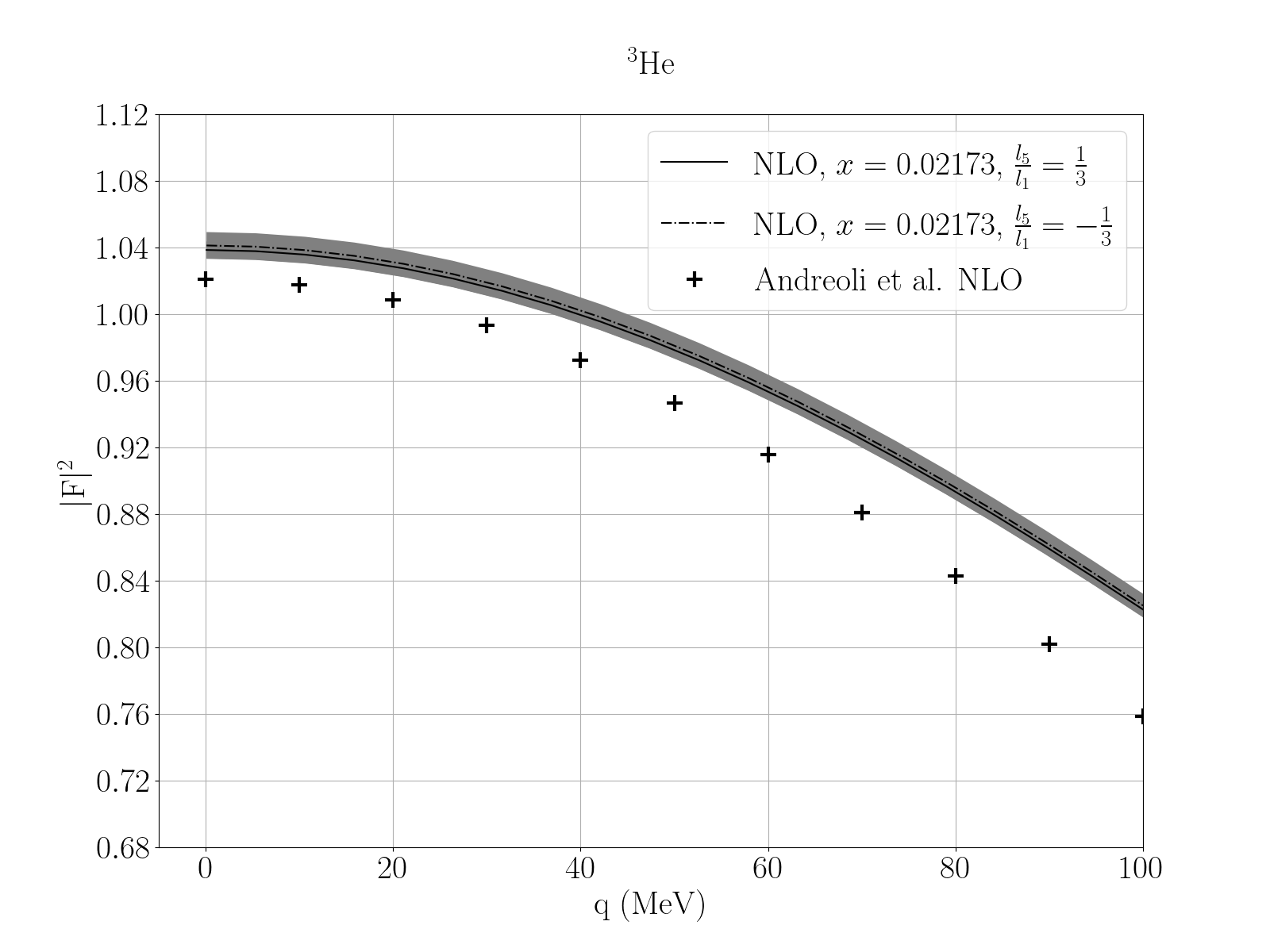}
    \end{subfigure}
    \caption{Spin-independent response functions of ${}^3$H and ${}^3$He. The crosses are the quantum Monte Carlo calculations of Ref.~\cite{Andreoli:2018etf} including only isoscalar contributions. The solid and dashed lines represent the response function including the large-\Nc constraint on $l^\chi_5$ with respect to $l^\chi_1$ with relative positive and minus signs, respectively. The gray bands represent $30\%$ uncertainty from the large-\Nc estimate.}
    \label{fig:dark_matter_response}
\end{figure}

Reference~\cite{Richardson:2021liq} considers the impact of large-\Nc constraints on the response functions of light nuclei interacting with heavy dark matter particles of spin 0 or spin 1/2 through a scalar current.
The relevant piece of the Lagrangian for spin-independent (SI) dark-matter-triton and helium-3 elastic scattering, which is the same for a heavy dark matter particle of either spin, is 
    \begin{align}     \label{eq:Lagrangian:dark_matter:two_nucleon_minimal}
        \calL_{\chi,NN} & =  C_{1, \chi NN}^{(\text{SI}, \, s)} \left( \chi^\dagger \chi \right)  \left( N^\dagger N \right) \left( N^\dagger N \right)  + C_{2, \chi NN}^{(\text{SI}, \, s)} \left( \chi^\dagger \chi \right)  \left( N^\dagger \sigma^i N \right) \left( N^\dagger \sigma^i N \right)  \nonumber \\
        &  + C_{1, \chi NN}^{(\text{SI}, \, v)} \left( \chi^\dagger \chi \right)  \left( N^\dagger \tau^3 N \right) \left( N^\dagger N \right) + \cdots  \, ,
    \end{align}
where $\chi$ is the external dark matter field and the superscripts $s$ and $v$ denote isoscalar and isovector operators, respectively.
According to the rules of Sec.~\ref{sec:aspects}, the LECs above scale as $C_{1, \chi NN}^{(\text{SI}, \, s)} \sim O(\Nc)$, $C_{2, \chi NN}^{(\text{SI}, \, s)} \sim O(1/\Nc)$, and $C_{1, \chi NN}^{(\text{SI}, \, v)} \sim O(1)$.
Working in terms of dibaryon fields the Lagrangian becomes
    \begin{equation}   \label{eq:triton:dibaryon_lagrangian}
        \begin{aligned}
            \calL_{\chi, ts} & = l^\chi_1  \chi^\dagger \chi  t_i^\dagger t_i + l^\chi_4 \left( \chi^\dagger \chi \right) s_a^\dagger s_a + l^\chi_5 i \epsilon^{3ab}  \left( \chi^\dagger \chi \right) s_a^\dagger s_b + \cdots \, ,
        \end{aligned}
    \end{equation}
where $t_i$ ($s_a$) is a spin-triplet (spin-singlet) dibaryon field.
The large-\Nc scalings of the LECs $l_i^\chi$ can be determined by matching the dark matter-deuteron scattering amplitudes of the two pictures.
This leads to
    \begin{equation}
        l_1^\chi \propto C_{1, \chi NN}^{(\text{SI}, \, s)} + C_{2, \chi NN}^{(\text{SI}, \, s)}, \quad l_4^\chi \propto C_{1, \chi NN}^{(\text{SI}, \, s)} -3 C_{2, \chi NN}^{(\text{SI}, \, s)}, \quad l_5^\chi \propto C_{1, \chi NN}^{(\text{SI}, \, v)} \, .
    \end{equation}
Therefore, $l_1^\chi$ and $l_4^\chi$ are both $O(\Nc)$ and should have the same sign while $l_5^\chi$ is relatively $1/\Nc$ suppressed.
By fixing $l_1^\chi$ to quantum Monte Carlo calculations for SI dark matter-deuteron scattering based on ChEFT currents \cite{Andreoli:2018etf}, Ref.~\cite{Richardson:2021liq} uses these scalings to  set the values of $l_4^\chi$, and $l_5^\chi$.
The resulting response functions are shown in Fig.~\ref{fig:dark_matter_response}.
The inclusion of only $l_1^\chi$ and $l_4^\chi$ is in agreement with the quantum Monte Carlo results, which are in the isospin limit, while the inclusion of the isovector term $l_5^\chi$ leads to a shift in the response functions by a few percent.

\section{THREE-NUCLEON FORCES}\label{sec:three}

The application of large-\Nc methods to interactions between more than two nucleons is still at an early stage. Reasons include the many possible spin-isospin and momentum structures in few-nucleon interactions, and the proliferation of contributing \NN interactions outside the two-nucleon COM frame. 
Reference \cite{Phillips:2013rsa} considers parity- and time-reversal-conserving three-nucleon operators from the large-\Nc perspective to provide some insight into their relative importance. 
As in the case of the \NN interactions, the operators can be written as combinations of the operators $\1,G^{ia},S^i,T^a$ for each nucleon, which are contracted with factors of momenta. 
As an example of the complexity of the three-nucleon sector, after taking into account all possible permutations of the three nucleons, Ref.~\cite{Phillips:2013rsa} finds 100 possible spin-isospin operators. 
These operators have to be combined with momenta to obtain rotational invariance.
The relevant momenta in the three-nucleon COM frame are 
\begin{equation}
    \vec{p} \equiv \vec{p}_1-\vec{p}_2\, , \quad \vec{q} \equiv \vec{p}_3 - (\vec{p}_1+\vec{p}_2)/2 \, ,
\end{equation}
where $\vec{p}_i$ $(i=1,2,3)$ is the momentum of the $i$-th incoming nucleon. Analogous definitions for $\vec{p}^{\,\prime}, \vec{q}^{\,\prime}$ hold for the outgoing nucleons. 
Similar to the \NN case, it is convenient to define the combinations 
\begin{equation}
    \vec{p}_\pm \equiv \vec{p}^{\,\prime}\pm \vec{p} \, , \quad \vec{q}_\pm \equiv \vec{q}^{\,\prime}\pm \vec{q}\, .
\end{equation}
However, unlike in the two-nucleon sector, $\vec{p}_+ \cdot \vec{p}_- \ne 0$; instead
\begin{equation}
    \vec{p}_+ \cdot \vec{p}_- = -\frac{4}{3}\vec{q}_+ \cdot \vec{q}_-.
\end{equation}
Factors of $\vec{p}_+$ and $\vec{q}_+$ both stem from relativistic corrections and are suppressed by factors of $1/m_N \sim \Nc^{-1}$; terms including these momenta correspond to nonlocal interactions. 
The \LONc terms therefore only contain $\vec{p}_-$ and $\vec{q}_-$ and are local.

Reference \cite{Phillips:2013rsa} finds 29  terms at leading order in the \Nc expansion. These consist of the identity operator and terms that are present in the traditional Fujita-Miyazawa three-nucleon force \cite{Fujita:1957zz}. 
Again in analogy to the symmetry-preserving two-nucleon sector, the expansion of the three-nucleon potential is in powers of $1/\Nc^2$. There are 645 terms that are suppressed by factors of $1/\Nc^2$ relative to the leading expressions, including 51 local operators. 
However, the authors of Ref.~\cite{Phillips:2013rsa} point out that they did not take into account permutation symmetry constraints, which may relate different operators structures, thus reducing the number of independent terms.

The existence of 80 local operators is confirmed in Ref.~\cite{Epelbaum:2014sea}. These terms can be generated from 20 operator structures by taking into account all possible permutations of the nucleon labels. 
As a check of the large-\Nc expectations, the authors consider a subset of pion-exchange diagrams contributing to the three-nucleon potential in the so-called equilateral triangle configuration, in which the distance between any nucleon pair is the same. 
For these particular contributions in this specific set-up, there is a clear hierarchy among the different  terms. 
The dominant terms are indeed those that are LO in the \Nc expansion. 
In addition, the authors also observe that several of the \LONc contributions are in fact negligible. Since the large-\Nc analysis only predicts the largest possible scaling with \Nc, this suppression does not contradict the large-\Nc results. 
While these results are promising, there are other contributions to the three-nucleon interactions that are not considered in the analysis of Ref.~\cite{Epelbaum:2014sea}. 
It would be interesting to study whether these contributions also adhere to the expected large-\Nc hierarchy and maybe provide significant contributions in the \LONc cases for which the contributions of Ref.~\cite{Epelbaum:2014sea} are small.

While many aspects of the large-\Nc analysis in the three-nucleon sector remain to be explored, there are certain similarities to the two-nucleon sector: there is a hierarchy among the allowed spin-isospin-momentum structures, with an expansion in powers of $1/\Nc^2$, and so far no substantial disagreement is found in a comparison to existing potentials.
However, similar caveats as in the two-nucleon sector apply. In particular, the ChEFT considered in Ref.~\cite{Epelbaum:2014sea} does not include explicit $\Delta$ degrees of freedom.

\section{ASPECTS OF NUCLEAR MATTER}\label{sec:nuclear}

So far the discussion in this review has focused on the interactions between two and three nucleons and the coupling to external currents. 
One of the goals of nuclear physics is to understand how the properties of complex nuclei and of nuclear matter arise from these interactions.
While in general the properties of the two- and three-nucleon interactions as extracted from few-nucleon experiments match the large-\Nc expectations, the large-\Nc expansion may not be sufficient to describe the much more complex physics of many-nucleon systems \cite{Gross:2011ve}. 

There are several open questions for systems with a large number of nucleons.
For example, the phase diagram of QCD in the large-\Nc limit looks different from that for $\Nc = 3$ \cite{McLerran:2007qj}. 
One point of view is that nuclear matter in the large-\Nc limit might be expected to form a crystal state in the large-\Nc limit \cite{Klebanov:1985qi}.
This expectation arises from a combination of aspects, including the baryon mass scaling as \Nc and the fact that the \NN interaction has a strong attractive contribution of order \Nc. 
Another point of view \cite{Bonanno:2011yr} claims that nuclear matter does not bind for large \Nc.
Also see Refs.~\cite{Torrieri:2010gz,Heinz:2011xq,Giacosa:2017mis} for related discussions.
Further, calculations based on Skyrmion models suggest that the binding energy should be of order $\Nc \Lambda_\text{QCD}$, in contrast to the relatively small binding energies observed in nature, see Ref.~\cite{Hidaka:2010ph} and references therein.
In addition, the authors of Ref.~\cite{Shuster:1999tn} conclude that  ``one should not expect the large \Nc limit to be of direct relevance for physics with $\Nc = 3$ at finite densities.''

These observations raise questions about whether the large-\Nc expansion can be usefully applied to understand the binding of heavier nuclei and the phases of nuclear matter.
One suggested resolution is that the nucleon axial coupling $g_A$ scales as $g_A\sim \Nc^0$ instead of the conventionally assumed scaling $g_A\sim \Nc$ \cite{Hidaka:2010ph,Kojo:2012hf,Chen:2017tgv}.
This modified scaling removes the large component in the \NN interaction, which may resolve the tension regarding the binding energies and the phase of nuclear matter.
This counting scheme was further explored in Ref.~\cite{Chen:2017tgv} in the context of hadronic effective theories of pion-nucleon and nucleon-nuceon scattering.
The main idea is that the large-\Nc counting rules need to be applied to scattering amplitudes, not to Feynman diagrams, and that a relativistic normalization is used for nucleon states.
The nonrelativistic reduction of the amplitudes results in additional factors of the nucleon mass $M\sim \Nc$, which modify the conventional scaling discussed in this review.
As a result, the scaling of the nucleon axial coupling  changes to $g_A\sim \Nc^0$. 
The author of Ref.~\cite{Chen:2017tgv} argues that analogous modifications also apply to other meson-nucleon couplings and that the modified scalings resolve the issues found in some multiple-meson-exchange contributions to \NN scattering, see Ref.~\cite{Belitsky:2002ni} and the discussion in Sec.~\ref{sec:mesonexchange}.

\section{CONCLUSIONS AND OUTLOOK}

In this review we have intended to provide specific guidance for establishing a hierarchy of two- and three-nucleon interactions based upon their behavior in the large-\Nc limit. 
This hierarchy is exhibited in the scaling behavior of  the LECs associated with each interaction. 
In Sec.~\ref{sec:aspects} we give the rules to follow, using the spin-isospin structure of the interaction, the source of momentum dependence, and the number of pions involved.  
We also emphasize the potential pitfalls of the approach. 
In particular, given the size of the expansion parameters (1/3 between isospin channels for $\Nc=3$) and the appearance of numerical factors that can obscure the hierarchy, we consider the ordering of interactions based upon large-\Nc as indicating trends rather than strict predictions.  
When data is available to determine the LECs, such as for \NN interactions in low partial waves, the large-\Nc analysis can provide an explanation of observed patterns in the relative sizes.
When no other information is available, such as in a BSM process and/or in interactions appearing at high orders in an EFT, the large-\Nc hierarchy can be used to inform allocation of resources for future experiments, lattice calculations, and many-body calculations. 

We provided examples of the procedure in two-nucleon strong interactions, weak interactions, and interactions with currents, including applications to neutrinoless double-beta decay and dark-matter-few-nucleon scattering response functions.   
For the strong interactions, we find that the large-\Nc constraints for EFT and phenomenological models give relative strengths that are in general agreement with \NN scattering data.
The success of this large-\Nc approach in the symmetry-conserving sector is encouraging and indicates that the approach may be useful for describing other aspects of few-nucleon systems. 
Of particular note is the identification of a variety of enhanced/accidental/approximate symmetries that may help illuminate trends in nuclear physics. 

In the symmetry-violating sector, application of large-\Nc constraints to PV interactions has so far not been in disagreement with available measurements, but in many cases these have large error bars. 
The PV interactions serve as an example of the usefulness of the large-\Nc formalism for two- and few-nucleon interactions when experimental constraints are scarce.
While the expansion is only in powers of $1/\Nc$ (as opposed to the expansion in $1/\Nc^2$ seen in the strong sector) and the available experimental information is not sufficient to provide a stringent test of the predicted large-\Nc hierarchy, application of the large-\Nc formalism has led to a reassessment of the previously prevailing ideas about which components of the PV\NN interactions are most important.
In particular, the finding that the isotensor component is \LONc has led to renewed interest in determining its value, either experimentally \cite{Gardner:2017xyl,Howell:2020nob} or through a lattice QCD calculation. 
As discussed in Ref.~\cite{Kurth:2015cvl}, the absence of contributions from disconnected diagrams in the isospin-symmetric limit makes the isotensor LEC an ideal target for lattice QCD studies of hadronic parity violation.
For TVPV interactions, the technology has been developed, but no data is yet available to test against. The least well developed sector is in TVPC interactions, which is not surprising considering that the effects are likely to be the smallest of those discussed here. 

Large-\Nc scaling rules applied to two-nucleon interactions with external currents finds agreement with experiment when available (magnetic currents), and agreement with other analyses when available (neutrinoless double-beta decay).
This provides further confidence that the rules will be helpful when applied to physics for which there is no other guidance, with dark matter detection presented as an example in this review.

The application of large-\Nc methods to three-nucleon interactions is less well developed and is a direction where further progress can be made. There are several  three-nucleon PV processes to be analyzed under the combined \eftnopi and large-\Nc expansions.  The situation for nuclear matter is even less clear, with the possibility that large-\Nc is not even a useful concept for analyzing the QCD phase diagram.

There exist a number of open questions and potential avenues for better understanding the impact of a large-\Nc analysis. 
For instance, while Ref.~\cite{Koerber:MS} discusses the consistency of using unitary transformations in the construction of the ChEFT \NN potential with the large-\Nc analysis, the application of the large-\Nc expansion to ChEFT interactions remains largely unexplored. 
Further, it may be that large-\Nc orderings of few-nucleon interactions get modified when they are embedded in a nucleus. 
Both lattice QCD (e.g., determining how EFT LECs scale in the presence of baryon density) and lattice EFT (e.g., whether enhancements occur in a multi-nucleon environment) may be helpful. 
There is also more to be understood about the role of the $\Delta$ in the procedure. 
Efforts are currently underway to explore the impact on large-\Nc relationships of including the $\Delta$ in intermediate states of few-nucleon potentials.  
There is also the interesting question of the apparent choices that can be made in the implementation of large-\Nc behavior.  
These include the scaling of typical nucleon momenta in a nucleus, the scaling of charges, and even the representation of the SU(\Nc) group used (see, e.g., Refs.~\cite{Cherman:2006iy,Cherman:2009fh,Lebed:2010um} for applications in the single-baryon sector).

\section*{ACKNOWLEDGMENTS}
This material is based upon work supported by the U.S.~Department of Energy, Office of Science, Office of Nuclear Physics,
under Award Numbers DE-SC0019647 (MRS) and DE-FG02-05ER41368 (TRR, RPS).
This work was supported in part by the Deutsche Forschungsgemeinschaft (DFG) through the Cluster of Excellence ``Precision Physics, Fundamental Interactions, and Structure of Matter'' (PRISMA${}^+$ EXC 2118/1) funded by the DFG within the German Excellence Strategy (Project ID 39083149) (TRR).
We are grateful to Aleksey Cherman, Zohreh Davoudi, Xincheng Lin, and Jared Vanasse for discussions.

\end{document}